\documentclass[12pt]{Definitions/academia}
\usepackage{hyperref}
\usepackage{cite}
\begin{document}
\articletype{Research article}
\proofstage{ Preprint}
\jName{Journal Name}

\title{Vulnerability of f(Q) gravity theory and a possible resolution.}
\author{$^{1}$Dalia Saha, $^{2*}$Abhik Kumar Sanyal}

\maketitle
\affiliation{$^{1}$Dept. of Physics, Jangipur College, Murshidabad, West Bengal, India - 742213,\& Dept. of Physics, University of Kalyani, Nadia, West Bengal, India - 741235.\\
$^{2}$ Dept. of Physics, Jangipur College, Murshidabad, West Bengal, India - 742213, \& Calcutta Institute of Theoretical Physics, Bignan Kutir, 4/1, Mohanbagan Lane, Kolkata, India - 700004.}

\noindent $^{*}$Correspondence:  Abhik Kumar Sanyal
\email{$^{*}$sanyal\_ak@yahoo.com}

\section*{Abstract}
Both the generalized teleparallel theories of gravity suffer from some serious problems. The strong coupling issue appearing as a consequence of extra degrees of freedom in the `generalized metric teleparallel gravity' theory, prompted to consider `generalized symmetric teleparallel gravity' theory (GSTG). Unfortunately, recent  perturbative analysis in the background of maximally symmetric space-time revealed that GSTG also suffers from the strong coupling issue and the ghost degrees of freedom. It has also been cognised that GSTG does not admit diffeomorphic invariance in general. Lately, it has been shown that except for the first, the other two connections associated with spatially flat Robertson-Walker metric do not even admit GSTG, while the first connection leads to an eerie Hamiltonian upon ensuing Dirac-Bergmann constraint analysis. Here we show that the only existing non-flat connection is also not viable in the same sense. Thus GSTG happens to be jeopardized. These problems do not showup in $f(R,Q)$ theory of gravity. Modified Dirac-Bergmann constraint analysis is deployed to formulate the phase-space structure. Quantization, probabilistic interpretation and semi-classical approximation connote that such a theory is well behaved in the context of early inflation, which has also been studied.
\keywords Generalized symmetric teleparallel gravity theory, {f(R,Q) gravity theory, Phase space structure, Inflation.}

\section{Introduction}\label{sec1}

More than a decade back teleparallel theory resurrected in it generalized form, $f(\mathrm{T})$ theory of gravity, also called the `generalized metric teleparallel gravity' (GMTG), to combat cosmic puzzle (recent accelerated expansion of the universe) without requiring dark energy, where `$\mathrm{T}$ is the torsion scalar'. However, GMTG carries additional degrees of freedom in comparison to `General Theory of Relativity' (GTR), which consequently give rise to the strong coupling issues \cite{T1,T2,T3,T4,T5,T6,T7,T8,T9,T10,T11,T12}. Strong coupling implies excessively large interactions between the extra degrees of freedoms (scalar fields) and matter. As a result the theory becomes complex due to the nonlinear nature of the equations, resulting in the breakdown of perturbation theory. This makes the theory unreliable in the early universe. Further, ghost degrees of freedom having negative norm states also appear \cite{TQ1}, which breaks unitarity. These issues prompted to consider `generalized symmetric teleparallel gravity' (GSTG), also known as `$f(Q)$ gravity' ($f_{,QQ} \ne 0$, where comma denotes derivative), where $Q$ is the `non-metricity scalar', built from the `non-metricity tensor' $Q_{\alpha\mu\nu} = \nabla_\alpha g_{\mu\nu} \ne 0$, where $g_{\mu\nu}$ is the metric tensor. In the last decade f(Q) gravity theory has been extensively studied from different perspectives, such as classical cosmology and late-stage of cosmic evolution \cite{1,2,3,4,5,6,7,8,9,10,11,12,13,14,15,16,17,18,19,20,21,22,22a,22b}, quantum cosmology \cite{23,24}, blackhole and wormhole physics \cite{25,26,27,28,29,30,31,32,33,34,35,36,37} and the study of compact objects \cite{38,39,40,41,42,43,44,45,46} etc. For a recent well-versed review on latest developments of $f(Q)$ gravity theory, it is suggestive to consult \cite{LH}.\\

However, recent findings reveal that this theory also runs from some serious pathologies, such as a large number of propagating modes, some of which are strongly coupled at high energy scale. This breaks down the linear perturbation theory around flat maximally symmetric background \cite{TQ1,Q1,Q2,Q3,Q4} potentially rendering the theory un-physical. Whatsoever, strong coupling issue may be alleviated in several ways. For example, in the higher-order perturbation analysis (which needless to say, is extremely difficult to compute) or restricting the number of propagating modes for some specific forms of $f(Q)$, or even identifying regimes where it is less severe, in the effective field theoretic approach etc. However, there exists other no-less serious problems. The Hamiltonian formulation of $f(Q)$ gravity not only brings about the ghosts but also fails to establish diffeomorphic invariance \cite{H1,H2} or even the `Dirac-Bergmann algorithm' itself, due to the fact that the inhomogeneous system of partial differential equations required to check consistency condition cannot be solved exactly \cite{H3}. These problems stem from the fact that the shift vector ($N_i$) and the auxiliary variable ($\Phi$) act as non-propagating dynamical variables and the auxiliary variable turns out to be the ghost. However, since the momentum canonical to the auxiliary variable appears as a function of the derivative of the shift vector, therefore in an appropriate and cosmologically viable minisuperspace model being devoid of the shift vector, the problems should not appear. In the spatially flat `Robertson-Walker (RW) metric' there exists three sets of non-vanishing connections \cite{Conn1, Conn2, Conn3}, yielding three different `non-metricity scalars', two of which are apparently nontrivial due to the presence of an undetermined functional parameter $\gamma(t)$. Very recently \cite{DA1} it is noticed that apart from the first one, the other two apparently non-trivial connections do not admit GSTG (for $f(Q)_{,QQ} \ne 0$). That is, apart from STEGR (Symmetric Teleparallel Equivalent of General Relativity), these two connections admit at-most non-minimally coupled scalar-vector-tensor theory of GSTG with regard to generalization \cite{Z}. However, associating a non-minimally coupled scalar field goes against the basic motivation of invoking teleparallel theories as alternatives to the dark energy. The first set of connections, for which the connection variation equation is trivially satisfied, does admit GSTG (for $f(Q)_{,QQ} \ne 0)$. However, it is found that the `scalar field' required to trigger inflation at the very `early universe' is also required to find a viable Friedmann-like $(a \propto \sqrt t)$ radiation dominated era. Unfortunately, the scalar jerks up and exhibits its very presence in the matter-dominated era too \cite{DA2}. In this process, the scalar field again acts as dark energy and hence the fundamental motivation of considering `teleparallel gravity' goes in vain. Further, in the course of formulating the `phase-space structure' for this connection (connection-1) in the absence of the shift vector, `diffeomorphic invariance' has although been established, the Hamiltonian turns out to be eerie and is not amenable \cite{DA1}. It may be mentioned that all these problems are artefact of maximally symmetric space-time, such as de-Sitter or spatially flat RW metric. Nonetheless, in the Robertson-Walker metric there also exists a non-flat ($k\ne 0$) connection, which has not been explored as such. Therefore here in particular, we study the only (left out) non-flat connection associated with `RW metric', henceforth will be dubbed as connection-4, and construct the `phase space structure'. Unfortunately, we shall see that the Hamiltonian here too turns out to be uncanny and is not maneuverable. Under such circumstances, it is viable to consider an even more generalized theory such as $f(R,Q)$ theory of gravity \cite{M}, where $R$ is the Ricci scalar built out of the `Levi-Civita connection' and $Q$ is the non-metricity scalar, which is our current concern. In such a theory, the higher-order terms appearing in the function $f(R)$ take control of higher degree terms \cite{KNRA} associated with $f(Q)$ theory, and hence the `ghost degree of freedom' does not appear.\\

The manuscript is organized in the following manner. In the next section, we briefly describe the geometry associated with `symmetric teleparallel gravity theory', write the field equations for $f(Q)$ gravity theory, briefly recapitulate the issues with all the three connections in spatially flat `RW metric'. Next in section 3, we consider the last and the only existing `non-flat $(k \ne 0)$' connection, find the field and the connection variation equations. It happens to admit $f(Q)$ GSTG, but only for negative curvature parameter $(k = -1)$ (note that $k=0$ leads to connection-3). Thereafter, we formulate the phase-space structure associated with connection-4 following ``Dirac-Bergmann constraint analysis''. As mentioned, the Hamiltonian yet again turns out to be eerie, which does not lead to a viable quantum theory. We therefore conclude that $f(Q)$ GSTG alone is not feasible in the early universe and a generalized  extension is necessary. In section 4, we therefore consider a generalized $f(R,Q)$ theory of gravity, and as an example choose a particular viable form, $f(R,Q) = \alpha_1 R + \beta R^2 + \alpha_2 Q + \gamma Q^2$, construct the `phase-space structure' following modified ``Dirac-Bergmann constrained analysis (DBCA)" \cite{A2}. Thereafter, we canonically quantize and perform semiclassical approximation. The oscillatory behaviour of the semi-classical wave function paves the way to consider classical field equations for the study of inflation according to `Hartle criteria' \cite{Hartle}. Finally, we associate a scalar field to drive inflation in the very early epoch of cosmic evolution, reduce the field equation to Friedmann-like ones in view of a hierarchy of Hubble flow parameters, study inflation for two different choices of the scalar field potential and compare the inflationary parameters with those constrained by the Planck data. Finally, we conclude in section 5. We also add an appendix to enunciate the effect of boundary terms on the phase-space structure.

\section{Geometric building block of generalized symmetric teleparallel gravity (GSTG):}

The general affine connection is a combination of the `Levi-Civita connection', the `Contortion tensor' and the `Disformation tensor', being expressed as,
\begin{eqnarray} \label{2.1} {\Gamma^\alpha}_{\mu\nu} = \{_\mu{^\alpha}_\nu\}  + {K^\alpha}_{\mu\nu} + {L^\alpha}_{\mu\nu},\end{eqnarray}
where, the Levi-Civita connection, the contortion tensor and the disformation tensor are given respectively as,
\begin{eqnarray}\label{2.2} \begin{split}&\{_\mu{^\alpha}_\nu\} = \frac{1}{2} g^{\alpha\lambda}\left(g_{\mu\lambda,\nu}+g_{\nu\lambda,\mu}-g_{\mu\nu,\lambda}\right),\\&
{K^\alpha}_{\mu\nu} = \frac{1}{2}g^{\alpha\lambda}\left(\mathrm {T}_{\lambda\mu\nu}-\mathrm{T}_{\mu\lambda\nu}-\mathrm{T}_{\nu\lambda\mu}\right),\\&{L^\alpha}_{\mu\nu} = \frac{1}{2}g^{\alpha\lambda}\left(Q_{\lambda\mu\nu}-Q_{\mu\lambda\nu}-Q_{\nu\lambda\mu}\right),\end{split}\end{eqnarray}
The Levi-Civita connection is expressed in terms of the derivatives of the metric tensor $(g_{\mu\nu})$, the contortion tensor is constructed as linear combination of torsion tensors $(\mathrm{T}_{\alpha\mu\nu})$ and the disformation tensor is an outcome of the non-metricity tensor $(Q_{\alpha\mu\nu})$ respectively. The general Riemann curvature tensor may be expressed in terms of the general affine connection as,
\begin{eqnarray} \label{Curv} {R^\alpha}_{\beta\mu\nu} = \partial_\mu{\Gamma^\alpha}_{\nu\beta} - \partial_\nu{\Gamma^\alpha}_{\mu\beta} + {\Gamma^\alpha}_{\mu\sigma}{\Gamma^\sigma}_{\nu\beta} - {\Gamma^\alpha}_{\nu\sigma}{\Gamma^\sigma}_{\mu\beta}.\end{eqnarray}
If the connection is not symmetric, the above curvature tensor denoted by $({{\bar R}^\alpha}_{\beta\mu\nu})$, is exclusively expressed in terms of the contortion tensor. On the other hand, for symmetric connection, it \eqref{Curv} is denoted by ${\tilde R^\alpha}_{\beta\mu\nu}$, which is expressed as the sum of the Riemann curvature tensors constructed out of Levi-Civita connection and a combination of the disformation tensors.\\

GMTG i.e., the generalized gravity with torsion $f(\mathrm{T})$ initially had encountered some serious problems with `Local Lorentz Invariance' (LLI) and strong coupling issues \cite{T12,TQ1,LLT}. This motivated yet another possible formulation of teleparallel gravity theory with non-metricity, viz., GSTG or more specifically $f(Q)$ in the flat $({\tilde{R}^\alpha}_{\;\;\beta\mu\nu}  = 0)$ and torsion-free $({\mathrm{T}_{\mu\nu}}^\lambda = 0)$ environment. Due to the vanishing curvature constraint in GSTG, the connection differs from the trivial connection by a general linear gauge transformation and becomes purely inertial. In the absence of torsion, the general affine connection \eqref{2.1} is expressed as ${\Gamma^\alpha}_{\mu\nu} = \{_\mu{^\alpha}_\nu\} + {L^\alpha}_{\mu\nu}$, in which the Christoffel connection mixes inertia with gravitation. As a result, the evolution of the universe is controlled both by ${L^\alpha}_{\mu\nu}$, the gravitational effect and by ${\Gamma^\alpha}_{\mu\nu}$, the  inertial effect. Inertial effect restores Lorentz covariance (LLI) and is parameterized by an arbitrary function $\xi^\alpha(x)$, called the St\"uckelberg fields. If connections vanish $({\Gamma^\alpha}_{\mu\nu} = 0)$, as for the coincidence gauge, the inertial effect disappears and LLI is lost. Given a metric tensor  $g_{\mu\nu}$, the only non-trivial object associated to the connection is the non-metricity tensor,
\begin{eqnarray}\label{2.11}Q_{\alpha\mu\nu} = \nabla_\alpha g_{\mu\nu} = g_{\mu\nu,\alpha} - g_{\nu\rho}{\Gamma^\rho}_{\mu\alpha} -g_{\rho\mu}{\Gamma^\rho}_{\nu\alpha} \ne 0,\end{eqnarray}
where $\nabla$ is the covariant derivative satisfying curvature-free and torsion-free conditions and $Q_{\alpha\mu\nu}$ is symmetric in the last two indices $\mu$ and $\nu$. Since the covariant derivative of the metric tensor does not vanish, therefore the term `non-metricity' is used to identify the tensor and the quadratic scalars of symmetric teleparallel gravity. Two different types of non-metricity vectors follows, viz.,
\begin{eqnarray}\label{2.12} Q_\alpha= g^{\mu\nu}Q_{\alpha\mu\nu} = {Q_{\alpha\nu}}^\nu;\;\;\;\tilde{Q}_\alpha= g^{\mu\nu}Q_{\mu\alpha\nu} = {Q_{\nu\alpha}}^\nu,\end{eqnarray}
in view of which one can construct the so-called super-potential tensor, which is essentially the non-metricity conjugate, as,
\begin{eqnarray}\label{2.13} P^{\alpha\mu\nu} = -{1\over 4}Q^{\alpha\mu\nu} +{1\over 2}Q^{(\mu\nu)\alpha} +{1\over 4}(Q^\alpha + \tilde{Q}^\alpha)g^{\mu\nu} -{1\over 4}g^{\alpha(\mu}Q^{\nu)}.\end{eqnarray}
Finally, the quadratic non-metricity scalar $Q$ is constructed from the curvature-free and torsion-free non-metricity tensor and is expressed as,
\begin{eqnarray} \label{2.14} Q = -\frac{1}{4}Q_{\alpha\mu\nu}Q^{\alpha\mu\nu} + \frac{1}{2}Q_{\alpha\mu\nu}Q^{\mu\nu\alpha} + \frac{1}{4}Q_\mu Q^\mu - \frac{1}{2}Q_\mu \tilde{Q}^\mu = R + \nabla_{\alpha}(Q^{\alpha}-{\tilde{Q}}^\alpha).\end{eqnarray}
Replacing the curvature scalar by the non-metricity scalar in the Einstein-Hilbert (EH) action, STEGR emerges. Nonetheless, for an arbitrary functional form $f(Q)$ the following action
\begin{eqnarray} \label{2.15} A_Q = A_{[g,\Gamma]} = \int f(Q)\sqrt{-g} d^4 x +\mathcal{S}_m,\end{eqnarray}
is usually called the generalized version of symmetric teleparallel theory of gravity (GSTG), which deviates from GTR to a great extent. Of-course, GSTG may also be constructed in the scalar-vector-tensor environment \cite{Z}, but as mentioned, it is associated with non-minimally coupled scalar field, which emerges in the form of dark energy in the matter-dominated era too. Since teleparallel theories were constructed as alternatives to the dark energy, so it goes against the primary motivation and is not therefore our current concern. Variation of the above action \eqref{2.15} with respect to the metric tensor $g^{\mu\nu}$ and the connection ${\Gamma^\alpha}_{\mu\nu}$, lead to the following field equations,
\begin{eqnarray} \label{2.16} \begin{split}&
\frac{2}{\sqrt{-g}} \nabla_\lambda (\sqrt{-g}f_{,Q}{P^\lambda}_{\mu\nu}) +\frac{1}{2}f g_{\mu\nu} + f_{,Q}(P_{\nu\rho\sigma} Q_\mu{}^{\rho\sigma} -2P_{\rho\sigma\mu}Q^{\rho\sigma}{}_\nu) = -\kappa T_{\mu\nu},\\& \nabla_\mu\nabla_\nu\left(\sqrt{-g} f_{,Q} {P^{\mu\nu}}_\lambda\right) = 0,\end{split}\end{eqnarray}
respectively. The metric variation equation \eqref{2.16} may also be expressed in the following covariant form \cite{Conn1,Conn2},
\begin{eqnarray}\label{2.16a} f_{,Q} {G}_{\mu\nu} + \frac{1}{2}g_{\mu\nu}\big(Q f_{,Q} - f(Q)\big)+ 2f_{,QQ} {\nabla}_\lambda Q P^\lambda{}_{\mu\nu} = -\kappa T_{\mu\nu},\end{eqnarray}
where, ${G}_{\mu\nu} = {R}_{\mu\nu} - \frac{1}{2} g_{\mu\nu} {R}$ is constructed from the Levi-Civita connection and $f_{,Q}$ as mentioned, stands for the derivative of $f(Q)$ with respect to $Q$. It is noteworthy that for $f_{,QQ} = 0$, the field equations \eqref{2.16a} are simply identical to those for GTR, while for constant non-metricity scalar $(Q = Q_0)$ one retrieves nothing but GTR along with a cosmological constant. In both the cases, the connection variation equation \eqref{2.16a} is trivially satisfied.\\

Now, if the Lie derivative of the connection as well as the space-time metric with respect to the isometry vector ($X$) associated with a space-time metric are set to vanish, i.e.,
\begin{eqnarray}\begin{split}\label{2.17} &\mathcal{L}_X{\Gamma^\mu}_{\alpha\beta} = X^\rho {\partial{\Gamma^\mu}_{\alpha\beta}\over \partial x^\rho} + {\Gamma^\mu}_{\rho\beta}{\partial X^\rho\over \partial x^\alpha} + {\Gamma^\mu}_{\alpha\rho}{\partial X^\rho\over \partial x^\beta} - {\Gamma^\rho}_{\alpha\beta}{\partial X^\mu\over \partial x^\rho} + \frac{\partial^2 X^\mu}{\partial x^\alpha\partial x\beta} = 0,\\&
\mathcal{L}_X {g}_{\mu\nu} = X^\rho \partial_\rho{g}_{\mu\nu} + \partial_\mu X^\rho {g}_{\rho\nu} + \partial_\nu X^\rho {g}_{\mu\rho}=0,
\end{split}\end{eqnarray}
where, the vector $X^\rho$ is the Killing vector associated with the symmetry of the space-time, then equations \eqref{2.17} in the curvature-free and torsion-free environment are utilized to find all possible connections involved with a space-time metric. Under the curvature-less and torsion-less conditions, the affine connection is expressed as,
\begin{eqnarray}\label{2.18} {\Gamma^\alpha}_{\mu\nu} = {\partial {x^\alpha}\over \partial\xi^\rho}\left({\partial^2 \xi^\rho\over \partial x^\mu\partial x^\nu}\right),\end{eqnarray}
where, the four scalar fields $\xi^\rho (x^\alpha)$ are arbitrary function of $x^\alpha$, and are called the St$\ddot{\mathrm{u}}$eckelberg fields associated with the diffeomorphism.  The above form of the affine connection ${\Gamma^\alpha}_{\mu\nu}$ preserves the Lorentz covariance in general, under the non-metricity framework. It may be mentioned that under a special choice $\xi^\rho (x) = x^\rho$ (this is possible only for those space-time metrics which can be expressed in cartesian co-ordinate system), the affine connection \eqref{2.18}, being the derivative of the delta function, vanishes. This particular special choice is called the coincidence gauge, for which covariant derivative leads to partial derivative, i.e.,
\begin{eqnarray}\label{2.19} Q_{\alpha\mu\nu} = \nabla_\alpha g_{\mu\nu} = g_{\mu\nu,\alpha},\end{eqnarray}
and the metric $(g_{\mu\nu})$ becomes the only independent variable. Although, it considerably simplifies all computations and therefore has largely been explored in the literature, nonetheless in the absence of the affine connection the theory is not LLI. This fundamental problem of physics may be circumvented taking up non-coincidence gauge and indeed \eqref{2.18} admits such non-trivial connections too. For example, in the following isotropic and homogeneous Robertson-Walker metric, which is our current concern,
\begin{eqnarray}\label{RW} ds^2 = -N^2 dt^2 + a^2(t)\left[{dr^2\over {1-kr^2}} + r^2(d\theta^2 + r^2 \sin^2 \theta d\phi^2)\right],\end{eqnarray}
where, $N(t)$ is the Lapse function, and $a(t)$ is the scale factor, there exists following six (three translational and three rotational) spatial Killing vectors $X^\rho$
\begin{eqnarray} \begin{split}& X_1 = \sin{\phi}\partial_\theta + {\cos{\phi}\over \tan{\theta}}\partial_\phi,\;\;X_2 = \cos{\phi}\partial_\theta + {\sin{\phi}\over \tan{\theta}}\partial_\phi,\;\;X_3 = \partial_\phi,\\&
X_4 = \sqrt{1-kr^2}\sin{\theta}\cos{\phi}\partial_r+{\sqrt{1-kr^2}\over r}\cos{\theta}\cos{\phi}\partial_\theta - {\sqrt{1-kr^2}\over r}{\sin{\phi}\over \sin{\theta}}\partial_\phi,\\&X_5 = \sqrt{1-kr^2}\sin{\theta}\sin{\phi}\partial_r+{\sqrt{1-kr^2}\over r}\cos{\theta}\sin{\phi}\partial_\theta - {\sqrt{1-kr^2}\over r}{\cos{\phi}\over \sin{\theta}}\partial_\phi,\\&X_6 = \sqrt{1-kr^2}\cos{\theta}\partial_r+{\sqrt{1-kr^2}\over r}\sin{\theta}\partial_\theta.\end{split}\end{eqnarray}
Correspondingly the non-vanishing affine connections may be computed as \cite{Conn1,Conn2,Conn3},
\begin{eqnarray}\label{Hoh1} \begin{split}& {\Gamma^0}_{00} = K_1,\;{\Gamma^0}_{11} = {K_2\over 1-kr^2},\;{\Gamma^0}_{22} = K_2r^2,\;{\Gamma^0}_{33}=K_2r^2\sin^2 \theta;\\&
{\Gamma^1}_{01} = K_3,\;{\Gamma^1}_{11}= {kr\over 1-kr^2},\;{\Gamma^1}_{22} = -r(1-kr^2),\;{\Gamma^1}_{33} = - r\sin^2{\theta}(1-kr^2);\\&
{\Gamma^2}_{02} = K_3,\;{\Gamma^2}_{12} = {\Gamma^2}_{21} ={1\over r}, {\Gamma^2}_{33} = -\sin{\theta}\cos{\theta};\\&
{\Gamma^3}_{02} = K_3,\;{\Gamma^3}_{13}={\Gamma^3}_{31} = {1\over r},\;{\Gamma^3}_{23} = {\Gamma^3}_{32} = \cot{\theta};\;\;\end{split}\end{eqnarray}
in view of equation \eqref{2.18}, where $K_i$ are functions of time. The above connections essentially gives four independent sets of connections, three of which correspond to the spatially flat $(k=0)$ case and one for non-flat $(k=\pm 1)$ case. In the following, we briefly demonstrate the outcome of the three flat cases in connection with GSTG. In the next section, we shall deal explicitly with the fourth connection.\\

\noindent
\textbf{Connection-1:}\\
The first set of independent connections, henceforth will be called connection-1, results under the conditions $k=0,~K_1 = \gamma(t),~K_2= K_3 = 0$. Consequently, the non-vanishing components of the affine connections presented in equation \eqref{Hoh1} are,
\begin{eqnarray}\label{Hoh11} \begin{split}& {\Gamma^0}_{00} = \gamma(t),\;\;{\Gamma^1}_{22} = -r,\;{\Gamma^1}_{33} = - r\sin^2{\theta},\;\; {\Gamma^2}_{12} ={1\over r},\\& {\Gamma^2}_{33} = -\sin{\theta}\cos{\theta},\;\;
{\Gamma^3}_{13} = {1\over r},\;{\Gamma^3}_{23} = \cot{\theta}.\end{split}\end{eqnarray}
As a result, one can compute the non-metricity scalar in view of \eqref{2.14}, which reads as \cite{Dima, Avik},
\begin{eqnarray} \label{Q1} Q = -6{\dot a^2\over N^2 a^2} =-{3\over 2} \left({\dot z^2\over N^2 z^2}\right).\end{eqnarray}
In the equation \eqref{Q1}, we have used the three-space metric $h_{ij} = a^2\delta_{ij} = z\delta_{ij}$ as the basic variable instead of the scale factor, since it leads to proper quantized version \cite{A11, A22}. Let us mention that the connection variation equation in this case is trivially satisfied. Now, if we consider the matter source to be a perfect barotropic fluid, then from the combination of the field equations one finds \cite{DA1},
\begin{eqnarray} \dot\rho +3H({\rho+p}) = -\left(3{H^2} +{1\over 2}Q\right)f_{,QQ}{\dot Q},\end{eqnarray}
where, $\rho$ is the matter density and $p$ stands for the thermodynamic pressure. Since, the energy conservation law $\dot\rho +3H({\rho+p}) = 0$ holds independently for thermodynamic fluid, so one ends up with the definition of the non-metricity scalar $Q$, which in turn admits the validity of energy condition for arbitrary fluid source. This of-course is a promising consequence of the connection-1 under consideration. It is therefore reasonable to explore the phase-space structure in the presence of a scalar field $\phi$, required to drive inflation in the very early universe, which has also been formulated earlier \cite{DA1}, resulting in
\begin{eqnarray}\begin{split} \label{H1} \mathrm{H} = N\Bigg[-\left(f-Q f_{,Q}\right)z^{3\over 2}-\frac{{p_z}^2\sqrt z}{6f_{,Q}}&+\frac{p_{\phi}^2}{2z^{3\over 2}}+ V(\phi)z^{3\over 2}\\& + \frac{\left(Q f_{,Q}-\frac{f}{2} - \frac{p_\phi^2}{4 z^{5\over 2}} +\frac{ V(\phi)}{2}\right)}{\big(zf_{,Q}^3-\frac{{p_z}^2 f_{,QQ}}{3}\big)} f_{,Q}\sqrt{z}~ p_{Q}p_z\Bigg]=  N\mathbb{H},\end{split}\end{eqnarray}
where the Hamiltonian is,
\begin{eqnarray} \begin{split}\label{H22} \mathbb{H} = -\left(f - Q f_{,Q}\right)z^{3\over 2}-\frac{{p_z}^2\sqrt z}{6f_{,Q}}&+\frac{p_{\phi}^2}{2z^{3\over 2}}+ V(\phi)z^{3\over 2}\\&+\frac{\left(Q f_{,Q}-\frac{f}{2} - \frac{p_\phi^2}{4 z^{5\over 2}} +\frac{ V(\phi)}{2}\right)}{\big(zf_{,Q}^3-\frac{{p_z}^2 f_{,QQ}}{3}\big)} f_{,Q}\sqrt{z}~p_{Q}p_z.\end{split}\end{eqnarray}
It is evident that diffeomorphic invariance (DI) has been established and there exists no ghost degree of freedom. Note that DI allows one to choose an appropriate gauge ($N=1$) as for GTR. Nonetheles as also noticed earlier while analyzing for GMTG theory \cite{MKA}, the presence of $p_z^2$ term in the denominator makes the Hamiltonian extremely unsettling. Further, the last term is peculiar, both the kinetic ($p_\phi$) and the potential ($V(\phi$) parts of the minimally coupled scalar field $\phi$ being now coupled with $f_{,Q}, p_Q, p_z$ in the Hamiltonian. Therefore, the Hamiltonian is not tractable. In a nut-shell, the first connection does not seem to be viable to study cosmological evolution.\\

\noindent
\textbf{Connection-2:}\\
Next, we consider the second independent set of connections (Connection-2) associated with an unknown time dependent parameter $\gamma(t)$. It is constructed under the choice, $k=0,~K_1 = \gamma(t) + {\dot \gamma\over \gamma};\;K_2 = 0;\;K_3 = \gamma$, resulting in the following non-vanishing components of the affine connections \eqref{Hoh1},
\begin{eqnarray}\label{Hoh2} \begin{split}& {\Gamma^0}_{00} = \gamma+{\dot\gamma\over\gamma},\; {\Gamma^1}_{01} = \gamma,\;{\Gamma^1}_{22} = -r,\;{\Gamma^1}_{33} = - r\sin^2{\theta};\\&
{\Gamma^2}_{02} = \gamma,\;{\Gamma^2}_{12} ={1\over r}, {\Gamma^2}_{33} = -\sin{\theta}\cos{\theta};\\&
{\Gamma^3}_{02} = \gamma,\;{\Gamma^3}_{13} = {1\over r},\;{\Gamma^3}_{23} = {\Gamma^3}_{32} = \cot{\theta}.\end{split}\end{eqnarray}
Consequently, the non-metricity scalar reads as \eqref{2.14},
\begin{eqnarray}\label{Q2}
Q =-{6\dot a^2\over N^2 a^2} + {3\gamma\over N^2}\left(3{\dot a\over a} - {\dot N\over N}\right) + {3\dot\gamma\over N^2} = -\frac{3{\dot z}^2}{2z^2N^2}+{3\gamma \over N^2}\left({3\dot z \over 2z}-{\dot N \over N}\right)+{3{\dot \gamma} \over N^2}.\end{eqnarray}
It may be mentioned that the connection variation equation \eqref{2.16} leads to
\begin{eqnarray} {d\over dt}\left[{a^3\dot Q \over N}f_{,QQ}\right] =0,\end{eqnarray}
and the fact that energy condition for barotropic perfect fluid holds independently, it turns out that either $\dot Q = 0$, or $f_{,QQ} = 0$ \cite{DA1}. On the contrary, the same (the connection variation equation) in view of the Lagrange multiplier technique implies \cite{DA1}
\begin{eqnarray}\label{CV2} {a^3\dot Q \over N}f_{,QQ} =0,\end{eqnarray}
resulting in either $\dot Q = 0$, or $f_{,QQ} = 0$ irrespective of the energy-momentum tensor conservation law. For both the cases STEGR emerges in view of the covariant form of the field equations \eqref{2.17}. Thus, connection-2 is also untenable in the context of GSTG.\\

\noindent
\textbf{Connection-3:}\\
There exists a third independent set of connections for spatially flat $(k=0)$ case under consideration, henceforth will be dubbed as connection-3. This is an outcome of the conditions, $K_1 = - {\dot \gamma\over \gamma};\;K_2 = \gamma;\;K_3 = 0$. The non-vanishing components of the affine connections are as follows \eqref{Hoh1},
\begin{eqnarray}\label{Hoh2} \begin{split}& {\Gamma^0}_{00} = -{\dot\gamma\over\gamma},\;\;\;\;{\Gamma^0}_{11} = \gamma;\;\;\;\; {\Gamma^0}_{22} = -\gamma r^2,\\&{\Gamma^0}_{33} =  \gamma r^2 \sin^2{\theta},\;{\Gamma^1}_{22} = -r,\;{\Gamma^1}_{33} = - r\sin^2{\theta}.\end{split}\end{eqnarray}
Consequently, in view of \eqref{2.14} the non-metricity  scalar reads as,
\begin{eqnarray}\label{Q3} Q  = -6{\dot a^2\over N^2 a^2}+{3\gamma\over a^2}\left({\dot a\over a} + {\dot N\over N}\right) + 3{\dot \gamma \over a^2} = -{3\over 2}{\dot z^2\over N^2 z^2} + {3\gamma\over z}\left({\dot z\over 2z} + {\dot N\over N}\right) + 3{\dot \gamma \over z^2}.\end{eqnarray}
As before, the energy conservation law for barotropic perfect fluid leads either to $\dot Q = 0$, or $f_{,QQ} = 0$, while the Lagrange multiplier technique reveals the same equation \eqref{CV2} leading to the same result without requiring energy conservation law \cite{DA1}. Clearly, connection-3 also turns out to be refutable in the context of GSTG.\\

{\textbf{Summary:} Summarily, symmetric teleparallel gravity is nothing but GTR apart from a divergence term and therefore to obtain richer structure than GTR, $f(Q)$ gravity theory $(f_{,QQ} \ne 0)$ was evoked, which we dub as Generalized Symmetric Teleparallel Gravity (GSTG). In the flat Robertson-Walker metric there exists three different independent sets connections. The problems associated with all these connections have been explored earlier \cite{DA1}. Here, we therefore briefly describe the issues. In a nut-shell, it is shown that the first connection is not viable since the Hamiltonian is not tractable, while the second and the third connections do not admit GSTG in the form $f(Q)$. It is therefore required to inspect the only remaining non-flat connection associated with the Robertson-Walker metric.\\

In the following section, we deal with the left out non-flat ($k \ne 0$) `connection-4' explicitly, find the field equations together with the connection variation equation in view of Lagrange multiplier technique. Thereafter, we administer the consequence of the connection variation equation to formulate the Hamiltonian following ``Dirac-Bergmann constraint analysis''.

\section{Field equations and phase-space structure for non-flat Connection-4:}

It has been mentioned in the introduction that $f(Q)$ GSTG ($f_{,QQ} \ne 0$) suffers from strong coupling problem and the linear perturbation theory breaks down around flat maximally symmetric background. In the previous section we have discussed some other earlier findings \cite{DA1}, which either exhibit lack of predictive power in the early stage of cosmological evolution (having an eerie Hamiltonian for connection-1) or does not admit GSTG at all  (connection-2 and connection-3). Thus, all the three connections associated with maximally symmetric space-time (spatially flat RW metric) are confutable. Fortunately, we have still one non-trivial connection for the non-maximally symmetric metric ($k \ne 0$) at hand. This is the fourth and the final set of connections and henceforth will be dubbed as connection-4. It is formed under the conditions, $K_1 = - {{k+\dot \gamma}\over \gamma};\;K_2 = {\gamma\over (1-kr^2)};\;K_3 = 0$, and as a result, the non-vanishing components of the affine connections are \eqref{Hoh1},
\begin{eqnarray}\label{Hoh2} \begin{split}& {\Gamma^0}_{00} = - {{k+\dot \gamma}\over \gamma},\;{\Gamma^0}_{11} = {\gamma\over (1-kr^2)^2},\;{\Gamma^0}_{22} = {\gamma\over (1-kr^2)}r^2,\;{\Gamma^0}_{33}={\gamma\over (1-kr^2)}r^2\sin^2 \theta;\\&
{\Gamma^1}_{11}= {kr\over 1-kr^2},\;{\Gamma^1}_{22} = -r(1-kr^2),\;{\Gamma^1}_{33} = - r\sin^2{\theta}(1-kr^2);\\&
{\Gamma^2}_{12} = {\Gamma^2}_{21} ={1\over r}, {\Gamma^2}_{33} = -\sin{\theta}\cos{\theta};\\&
{\Gamma^3}_{13}={\Gamma^3}_{31} = {1\over r},\;{\Gamma^3}_{23} = {\Gamma^3}_{32} = \cot{\theta} .\end{split}\end{eqnarray}
Consequently, in view of \eqref{2.14} the non-metricity  scalar reads as,

\begin{eqnarray} \label{AQ4}Q =  -6{\dot a^2\over N^2 a^2}+{3\gamma\over a^2}\left({\dot a\over a} + {\dot N\over N}\right) + 3{\dot \gamma \over a^2} +k \left[{6\over a^2} + {3\over \gamma N^2}\left({\dot N\over N} + {\dot\gamma\over \gamma} -3{\dot a \over a}\right)\right].\end{eqnarray}
Now, treating the above non-metricity scalar \eqref{AQ4} as a constraint and introducing it through a Lagrange multiplier $\lambda$ into the action \eqref{2.15}, we recast the action as (we shall shift over to $z = a^2$ later in time),
\begin{eqnarray}\begin{split} \label{AC2}A = &\int \Bigg[f(Q) -\lambda\Bigg(Q+6{\dot a^2\over N^2 a^2}-{3\gamma\over a^2}\Big({\dot a\over a} + {\dot N\over N}\Big) - 3{\dot \gamma \over a^2}\\&
-k \left\{{6\over a^2} + {3\over \gamma N^2}\Big({\dot N\over N} + {\dot\gamma\over \gamma} -3{\dot a \over a}\Big)\right\}\Bigg)- \rho_0 a^{-3(\omega+1)} \Bigg]Na^3d t.\end{split}\end{eqnarray}
In the above, we have also taken a barotropic fluid source into account. Varying the action with respect to $Q$ one gets, $\lambda = {df\over d Q} = f_{,Q} $. Substituting it back, the action \eqref{AC2} may finally be expressed as,
\begin{eqnarray}\begin{split}\label{A2} A =& \int \Bigg[fa^3-Q f_{,Q}a^3-6f_{,Q}{a\dot{a}^2\over N^2}+ {3\gamma a f_{,Q}}\left({\dot a\over a} + {\dot N\over N}\right)+{3f_{,Q}a\dot{\gamma}}\\&+ kf_{,Q}\Big[6a+\frac{3\dot Na^3}{\gamma N^3}+\frac{3\dot \gamma a^3}{\gamma^2N^2}-\frac{9\dot a a^2}{\gamma N^2} \Big]- \rho_0 a^{-3\omega}\Bigg]Ndt.\end{split}\end{eqnarray}
Now, under the variations of $N, a, Q$ and $\gamma$ respectively, the following field equations are obtained.

\begin{eqnarray} \label{F4.1}{1\over 2} f + \left({3\mathcal{H}^2} -{1\over 2}Q\right)f_{,Q} - 3\gamma{\dot Q f_{,QQ}\over 2a^2}+3k \left({f_{,Q}\over a^2} -{\dot Q f_{,QQ}\over 2\gamma N^2} \right)= \rho,\end{eqnarray}
\begin{eqnarray}\label{F4.2}{1\over 2} f + \left({3\mathcal{H}^2} -{1\over 2}Q\right)f_{,Q} + {2\over N}{d\over dt} \left({\mathcal{H}} f_{,Q}\right) - \gamma {\dot Q f_{,QQ}\over 2a^2}+k \left({f_{,Q}\over a^2} -{3\dot Q f_{,QQ}\over 2\gamma N^2} \right)= -p,\end{eqnarray}
\begin{eqnarray} \label{F4.3}Q = -{6\mathcal{H}^2}+ {3\gamma N\over a^2}\left(\mathcal{H} + {\dot N \over N^2}\right) + {3\dot\gamma\over a^2}+k \left[{6\over a^2}+{3\over N\gamma } \left({\dot N\over N^2} + {\dot \gamma \over N\gamma} - 3\mathcal{H}\right)\right]\end{eqnarray}
\begin{eqnarray}\label{F4.4}\dot{Q}f_{,QQ}\left(N+{ka^2\over \gamma^2 N}\right)=0,\end{eqnarray}
where, $\mathcal{H} = {\dot a \over aN}$. Note that, the third ($Q$ variation) equation \eqref{F4.3} gives back the definition of the non-metricity scalar \eqref{AQ4} itself, while from the $\gamma$ variation equation \eqref{F4.4} it may be concluded, as in the cases of previous connections, that either $\dot Q = 0$ or $f_{,QQ} = 0$, none of which results in GSTG. However, here exists yet another intriguing option, viz., $\left(N+{ka^2\over \gamma^2 N}\right)=0$, for which GSTG in the form of $f(Q)$ gravity theory ($f_{,QQ} \ne 0$) is admissible. Clearly, connection-4 needs study since it appears to yield richer structure than GTR at least for the last case. Now, for the last choice, the connection variation equation thus implies,

\begin{eqnarray} \label{4g} \gamma^2 = -{ka^2\over N^2}; \;\;\;\mathrm{or},\;\;\;\gamma = \pm{a\over N},\end{eqnarray}
since only $k = -1$ is admissible for real function $\gamma(t)$. Thus, connection-4 is the only one that may have some richer structure than GTR, as mentioned. Note that, in view of equation \eqref{4g}, the connection variation equation \eqref{F4.4} is identically satisfied and in the process, the non-metricity scalar \eqref{F4.3} takes following two different forms, viz.,

\begin{eqnarray}\label{Q} Q = -6\mathcal{H}^2 \pm 12{\mathcal{H}\over a} - {6\over a^2} = -6\left(\mathcal{H} \mp {1\over a}\right)^2 = -6\left({\dot a\over N a} \mp {1\over a}\right)^2,\end{eqnarray}
corresponding to two different signs of $\gamma$ appearing in \eqref{4g}. As demonstrated in \cite{DA1}, it is possible to construct a scalar-tensor theory of GSTG in view of an auxiliary scalar field $\Phi$ in the following form,
\begin{eqnarray}\label{Action} A_{[Q,\Phi]} = \int [f'(\Phi) Q + f(\Phi) - \Phi f'(\Phi)]\sqrt{-g} d^4x + \mathcal{S}_m,\end{eqnarray}
where, prime  denotes derivative with respect to the dynamical auxiliary variable $\Phi$. Note that the above action \eqref{Action} is \eqref{2.15} in disguise, which is apparent taking the variation of action \eqref{Action} with respect to $\Phi$, whence one finds $\Phi = Q$. Therefore, the scalar-tensor action \eqref{Action} of GSTG may be expressed as,
\begin{eqnarray}\label{phiA1} A=\int\left[ f'(\Phi)\left\{R + \nabla_{\alpha}(Q^{\alpha}-{\tilde{Q}}^\alpha)\right\}+f(\Phi)-\Phi f'(\Phi)+L_m \right]\sqrt{-g} d^4x.\end{eqnarray}
In terms of the function $z = a^2$, the action \eqref{phiA1} reads as,
\begin{eqnarray}\begin{split}\label{phiA2.2} A=\int\Bigg[\left\{f(\Phi)- \Phi f'(\Phi)\right\}{Nz^{3\over 2}}-&6\left({{\dot z}^2\over {4N\sqrt z}}\mp \dot z+N{\sqrt z}\right)f'(\Phi)\\&+\left({1\over 2N^2} \dot\phi^2 - V(\phi)\right) Nz^{3\over 2}\Bigg]dt,\end{split}\end{eqnarray}
where, instead of a perfect fluid, we introduce the scalar field $\phi$ for driving inflation in the very early universe. In view of the above equation \eqref{phiA2.2}, the canonical momenta are found as,
\begin{eqnarray} \label{Ap1} p_{z}={\partial L\over \partial{\dot z}}=-{3{\dot z}\over {N\sqrt z}}f'(\Phi)\pm 6f'(\Phi),\hspace{0.1 in} p_{\phi}={\partial L\over \partial{\dot \phi}}={\dot\phi z^{3\over 2}\over N}, \hspace{0.1 in} p_{\Phi}={\partial L\over \partial{\dot \Phi}} = 0, \hspace{0.1 in} p_{N}={\partial L\over \partial{\dot N}} = 0.\end{eqnarray}
Therefore, one encounters two primary constraints associated with the theory, viz.,
\begin{eqnarray}\label{Aconstraints}\lambda_1 = p_{\Phi} \approx 0, \hspace{0.2 in}\lambda_2 = p_N \approx 0,\end{eqnarray}
which are second class constraints. However take note that, the associated constraint $p_N \approx 0$ vanishes strongly since none of the momenta densities contain $\dot N$. The lapse function therefore is non-dynamical and can be safely ignored. But, it is not legitimate to ignore $\Phi$ by employing the same argument since, it has been demonstrated \cite{Hu} that in the conformal transformed action $\dot \sigma$ appears, which is the conformal field that replaces $\Phi$. Therefore, the auxiliary field should tacitly assumed to be dynamical as considered in \cite{DA1}. Hence, it is only required to introduce the constraint associated with $p_{\Phi}$ through Lagrange multipliers $u_{1}$ into the following constrained Hamiltonian as,
\begin{eqnarray} \label{AHC} \mathrm{H}_c =\sum_i p_i \dot q_i - L = \dot z p_z + \dot \phi p_\phi +\dot \Phi p_\Phi+ \dot N p_N - L,\end{eqnarray}
and the primary hamiltonian may be expressed as,
\begin{eqnarray}\label{AHP}\begin{split} \mathrm{H}_{p} &= \mathrm{H}_c + u_1 p_{\Phi}\\&
 = -\frac{{p_z}^2N\sqrt z}{6f'}\pm 2N\sqrt zp_z-\left(f-\Phi f'\right)Nz^{3\over 2}+\frac{Np_{\phi}^2}{2z^{3\over 2}}+ NV(\phi)z^{3\over 2}+ u_1 p_\Phi. \end{split}\end{eqnarray}
Note that the Poisson bracket $\{z,p_{z}\}=\{\phi, p_{\phi}\} = \{\Phi,p_{\Phi}\}=1$ holds. Now, constraint should remain preserved in time, which is exhibited through the following poisson's bracket,
\begin{eqnarray}{\dot\lambda_{1}}=\left\{ \lambda_{1}, \mathrm{H}_{p}\right\}= -Nf''\left[\Phi z^{3\over 2}+\frac{{p_z}^2\sqrt z}{6f'^2}\right]+\sum^2_{i=1}\lambda_i\{\lambda_1,u_i\}\approx 0.\end{eqnarray}
Hence, the Lagrange multiplier $u_1$ remains obscure and the primary constraint $\lambda_1$ results in a secondary constraint, viz.,
\begin{eqnarray} \label{Aconst2}\lambda_2= -{Nf''}\left[\Phi z^{3\over 2}+\frac{{p_z}^2\sqrt z}{6f'^2}\right]\approx 0,\end{eqnarray}
The above constraint is required to be introduced into the primary Hamiltonian \eqref{AHP}, through yet another Lagrange multiplier. The modified primary Hamiltonian therefore reads as,
\begin{eqnarray}\begin{split}\label{AHc1} \mathrm{H}_{p1}= -\left(f-\Phi f'\right)Nz^{3\over 2}-\frac{N{p_z}^2\sqrt z}{6f'}\pm 2N\sqrt zp_z&+\frac{Np_{\phi}^2}{2z^{3\over 2}}+ NV(\phi)z^{3\over 2}\\&+ u_{1}p_{\Phi}-u_2Nf''\left[\Phi z^{3\over 2}+\frac{{p_z}^2\sqrt z}{6f'^2}\right],\end{split}\end{eqnarray}
where $u_2$ is yet another Lagrange multiplier. However at this end, to preserve the constraints over time, we compute the Poisson brackets yet again, with the modified primary Hamiltonian to find,
\begin{eqnarray}\label{APoisson1.1}\begin{split}&{\dot\lambda_{1}}=\left\{ \lambda_{1}, \mathrm{H}_{p1}\right\}=-Nf''\left[\Phi z^{3\over 2}+\frac{{p_z}^2\sqrt z}{6f'^2}\right] \\& \hspace{2.0 cm}+ u_2 N\left[f''\big(z^{3\over 2}-\frac{{p_z}^2\sqrt z f''}{3f'^3}\big)+f'''\big(\Phi z^{3\over 2}+\frac{{p_z}^2\sqrt z}{6f'^2}\big)\right] +\sum^2_{i=1}\lambda_i\{\lambda_1,u_i\}\approx 0,\end{split}\end{eqnarray}
\begin{eqnarray}\label{APoisson1.2}\begin{split}&{\dot\lambda_{2}}=\left\{ \lambda_{2}, \mathrm{H}_{p1}\right\}=N^2f''\left(\frac{\Phi z{p_z}}{f'}-\frac{z{p_z}f}{2f'^2}- \frac{z p_z p_\phi^2}{4f'^2 z^{5\over 2}} +\frac{z p_z V(\phi}{2f'^2}\right)\\& \hspace{2.0 cm}-u_1N{\left[f''\big(z^{3\over 2}-\frac{{p_z}^2{\sqrt z}f''}{3f'^3}\big)+f'''\big(\Phi z^{3\over 2}+\frac{{p_z}^2\sqrt z }{6f'^2}\big)\right]}+\sum^2_{i=1}\lambda_i\{\lambda_2,u_i\}\approx 0.\end{split}\end{eqnarray}
The above Poisson brackets reveals the following forms of the two Lagrange multipliers $u_1$ and $u_2$,
\begin{eqnarray}\label{ALagmult2}\begin{split}& u_1=\frac{Nf'' z p_z\left(\frac{\Phi}{f'}-\frac{f}{2f'^2} - \frac{p_\phi^2}{4f'^2 z^{5\over 2}} +\frac{ V(\phi)}{2f'^2}\right)}{\left[f''\big(z^{3\over 2}-\frac{{p_z}^2{\sqrt z}f''}{3f'^3}\big)+f'''\big(\Phi z^{3\over 2}+\frac{{p_z}^2\sqrt z }{6f'^2}\big)\right]} \\&u_2=\frac{f''\left(\Phi z^{3\over 2}+\frac{{p_z}^2\sqrt z }{6f'^2}\right)}{\left[f''\big(z^{3\over 2}-\frac{{p_z}^2\sqrt z f''}{3f'^3}\big)+f'''\big(\Phi z^{3\over 2}+\frac{{p_z}^2\sqrt z }{6f'^2}\big)\right]}.\end{split}\end{eqnarray}
Now, replacing the auxiliary variable $\Phi$ by $Q$, so that $f = f(Q)$, and using the definition of $p_z$ \eqref{Ap1}, it is possible to find the following relation,
\begin{eqnarray}\begin{split} &f_{,QQ}\left[Q z^{3\over 2} + {\sqrt z\over 6 f_{,Q}^2}\left(\frac{9{\dot z}^2f_{,Q}^2}{N^2z}+36f_{,Q}^2\mp\frac{36\dot zf_{,Q}^2}{N\sqrt z}\right)\right] \\& \hspace{4.0 cm}= f_{,QQ}z^{3\over 2}\left[Q + \left({3\dot z^2\over 2N^2 z^2}+{6\over z} \mp {6\dot z\over Nz^{3\over 2}} \right)\right] = 0,\end{split}\end{eqnarray}
where the definition of $Q$ is used. Clearly, the Lagrange multiplier $u_1$ is simplified while the Lagrange multiplier $u_2$ vanishes identically. Hence, one finally ends up with,
\begin{eqnarray}\label{Lagmult3}u_1 = \frac{\left(\frac{Q}{f_{,Q}}-\frac{f}{2f_{,Q}^2} - \frac{p_\phi^2}{4f_{,Q}^2 z^{5\over 2}} +\frac{ V(\phi)}{2f_{,Q}^2}\right)}{\big(z-\frac{{p_z}^2 f_{,QQ}}{3f_{,Q}^3}\big)} N \sqrt{z} p_z = \frac{\left(Q f_{,Q}-\frac{f}{2} - \frac{p_\phi^2}{4 z^{5\over 2}} +\frac{ V(\phi)}{2}\right)}{\big(zf_{,Q}^3-\frac{{p_z}^2 f_{,QQ}}{3}\big)} N f_{,Q}\sqrt{z} p_z,\end{eqnarray}
remembering the fact that for generalized teleparallel gravity theories $f_{,QQ} \ne 0$. In view of the above form of Lagrange multiplier $u_1$ \eqref{Lagmult3}, the Hamiltonian being free from the constraints may finally be expressed as,
\begin{eqnarray}\begin{split} \label{H1}\mathrm{H} = N \mathbb{H} = 0 = N\Bigg[-\left(f-Q f_{,Q}\right)z^{3\over 2}&\pm 2\sqrt zp_z -\frac{{p_z}^2\sqrt z}{6f_{,Q}}+\frac{p_{\phi}^2}{2z^{3\over 2}}+ V(\phi)z^{3\over 2}\\&+ \frac{\left(Q f_{,Q}-\frac{f}{2} - \frac{p_\phi^2}{4 z^{5\over 2}} +\frac{ V(\phi)}{2}\right)}{\big(zf_{,Q}^3-\frac{{p_z}^2 f_{,QQ}}{3}\big)} f_{,Q}\sqrt{z}~ p_{Q}p_z\Bigg],\end{split}\end{eqnarray}
where the constrained Hamiltonian is,
\begin{eqnarray} \begin{split}\label{H24}\mathbb{H} = -\left(f- Q f_{,Q}\right)z^{3\over 2}\pm 2\sqrt z p_z&-\frac{{p_z}^2\sqrt z}{6f_{,Q}}+\frac{p_{\phi}^2}{2z^{3\over 2}}+ V(\phi)z^{3\over 2}\\&+\frac{\left(Q f_{,Q}-\frac{f}{2} - \frac{p_\phi^2}{4 z^{5\over 2}} +\frac{ V(\phi)}{2}\right)}{\big(zf_{,Q}^3-\frac{{p_z}^2 f_{,QQ}}{3}\big)} f_{,Q}\sqrt{z}~p_{Q}p_z.\end{split}\end{eqnarray}
Diffeomorphic invariance is thus established and there exists no ghost degree of freedom. Surprisingly, the Hamiltonian takes almost the same form as \eqref{H22} found for connection-1 apart from the additional linear term $\pm 2\sqrt z p_z$, which is an outcome of the non-flat curvature parameter $k = -1$. Thus, connection-4 also suffers from the same pathologies discussed for connection-1 and is not tractable.\\

{\textbf{Summary:} In a nut-shell, along with all the three flat connections considered in the previous section 2, the connection-4 also does not seem to be feasible to study cosmological evolution. It is worth-mentioning that, the same form of the Hamiltonian may even be obtained without seeking auxiliary variable $\Phi$. It is therefore conclusive that GSTG in the form $f(Q)$ alone ($f_{,QQ} \ne 0$) is untenable and all the research oriented with GSTG theory over last one decade are in vain. Hence, it is rationale to search for a viable alternative that might be able to unfurl cosmic history without the need of dark energy, which is our next motive.

\section{$f(R,Q)$ theory of gravity:}

GSTG ($f_{,QQ} \ne 0$) not only contested dark-energy as an alternative successfully for more than a decade, but also can explain scalar-driven inflation along with other gravitational phenomena (Blackhole, Wormhole etc.), appreciably. The problems, as mentioned appeared in connection with the Hamiltonian formulation, since the theory does not admit diffeomorphic invariance. Nonetheless, this problem may be alleviated in the absence of the shift vector ($N^i$), i.e., for the space-time of cosmological interest, such as RW metric. However, it is found that for maximally symmetric space-time (such as de-Sitter, or flat RW metric), it fails to provide a viable linear perturbation. This means it is not possible to explain structure formation and the formation of CMB (Cosmic microwave background). Nevertheless, this problem may also be alleviated taking into account higher-order perturbation and also for non-flat RW space-time. Even so, it has recently been found that the two out of the three sets of  connections for flat RW space-time do not allow generalization of $f(Q)$ theory with $f_{,QQ} \ne 0$ and $Q \ne Q_0$, and therefore are equivalent to GTR in disguise, while the first one does not provide a quantum description of the theory, due to the eerie structure of the Hamiltonian. So only situation requires to explore is the non-flat RW space-time, which is non-maximally symmetric and therefore might pass through the perturbative analysis. Howbeit, in the previous section it is shown that this connection admits negative curvature ($k = -1$) and its Hamiltonian counterpart is again eerie. It may be mentioned that inflation is usually studied in view of classical field equations, while it is essentially a quantum theory of perturbation. Thus, unless a viable quantum theory is formulated, whose semi-classical wave-function is oscillatory, all the theoretical inflationary data of GSTG although agrees with the observational Planck's data, these results are not reliable.\\

Since GSTG theory ($f_{,QQ} \ne 0$) turns out to be untenable, we are therefore prompted to search for an alternative in the form $f(R,Q)$ gravity theory, in concurrence with our earlier proposal of $f(R, \mathrm{T})$ gravity theory, originally introduced to get rid of the issue of `branched Hamiltonian' \cite{MKA}. Let us mention that whenever a metric is chosen, curvature can be formulated out of it using Levi-Civita connection. Note that, the equation \eqref{2.1} for general affine connection is the sum of `Levi-Civita connection' and two tensors. Difference of two connections is a tensor. Therefore, if one of the tensors (contortion/disformation) is set to vanish, then the other has to be the difference between the affine connection and the `Levi-Civita connection'. This Levi-Civita connection, which forms the curvature of space-time, must not sit idle without playing any role. Thus, there is no reason to ignore the presence of curvature and a much more general form should be $f(R,Q)$. To explore the role of $f(R,Q)$, we make a particular physically reasonable choice viz., $f(R,Q) = \alpha_1R+ \beta R^2 + \alpha_2 Q + \gamma Q^2$, to start with, in this article (The constant coefficient $\gamma$ used here must not be confused with the $\gamma(t)$ that appeared in the expression of Q in connection 4). The second term of this form being an outcome of all the attempts to quantize gravity starting from Stell's work \cite{Stelle}, plays a dominant and vital role in the strong gravity regime (early universe and near black holes). Further, this term being associated with higher-order terms, controls the effect of the higher-degree terms appearing from $Q^2$ term and consequently is responsible to furnish Schr\"odinger-like equation upon quantization (as we shall see shortly), in which a physical parameter viz., the proper volume plays the role of time parameter. The linear terms (first and third) would provide a viable radiation as well as early matter dominated eras, in agreement with the standard Friedmann-Lema$\hat{\mathrm{i}}$tre-Robertson-Walker (FLRW) model. The $Q^2$ term, as we shall see, plays a dominant role in the inflationary regime. Note that, late time accelerated expansion requires yet another term, which may be in the form $Q^n$, where $n < {1\over 2}$, so that it can play a dominant role at the late-universe only, rendering recent accelerated expansion. At the early universe, this term remains subdominant and therefore, for the sake of simplicity, we drop this term from our current analysis in connection with the early universe. The action therefore is given by,
\begin{eqnarray} \label{A1.0} A = \int f(R,Q) \sqrt{-g} d^4x = \int (\alpha_1R+\beta R^2 + \alpha_2 Q + \gamma Q^2)\sqrt{-g} d^4x.\end{eqnarray}
where, the Ricci scalar in the spatially flat Robertson-Walker metric \eqref{RW} reads as,
\begin{eqnarray} \label{R} R = {6\over N^2}\left({\ddot a\over a}+{\dot a^2\over a^2}-{\dot N\dot a\over N a}\right) = {6\over N^2}\left({\ddot z\over 2 z}-{\dot N\dot z\over 2N z}\right).\end{eqnarray}
Any of the three formalisms, viz., Modified Horowitz' Formalism (MHF) \cite{A11, A22}, Dirac-Bergmann constrained analysis (DBCA) \cite{A2}, or Modified Buchbinder-Lyakhovich formalism (MBLF) \cite{BL, BL1, BL2, BL3} can be called for canonical formulation of higher-order theory of gravity, provided divergent terms appearing in the action are taken care of a priori. However, here we shall consider DBCA in its modified form (taking care of the divergent terms a priori) to construct the  Hamiltonian.

\subsection{Modified Dirac-Bergmann constraint analysis:}

In Dirac-Bergmann constraint analysis, total derivative terms are not taken care of, resulting in a different Hamiltonian, whose quantum counterpart is unfeasible (see appendix). The classical Hamiltonian although is related to the one constructed out of Modified Horowitz's formalism (MHF) or Modified Buchbinder Lyakhovich formalism (MBLF) under canonical transformation (see appendix), such transformation cannot be extended to the quantum domain due to non-linearity. We therefore modify the formalism by taking care of the total derivative terms and call it Modified Dirac-Bergmann constraint analysis (MDBA). Since in MDBA, it is customary to fix $h_{ij}$ and $K_{ij}$ at the boundary, the total derivative terms automatically vanish at end points and therefore there is no need to supplement the action with surface terms. Hence, in MDBA we start from the action (\ref{A1.0}), substitute the expressions for $R$ \eqref{R} and $Q$ \eqref{Q1} for connection-1, integrate by parts to eliminate the divergent term, so that the corresponding point Lagrangian may finally be expressed in the form,
\begin{center}\begin{eqnarray} \label{DA1}\begin{split} L&= -\frac{3\alpha_1\dot z^2}{2N\sqrt z}+\frac{9\beta}{\sqrt z}\bigg{(}\frac{\ddot z^2}{N^3}-\frac{2\dot z \ddot z \dot N}{N^4}+\frac{\dot z^2\dot N^2}{N^5} \bigg{)}-\frac{3\alpha_2\dot z^2}{2N\sqrt z}+\frac{9\gamma \dot z^4}{4N^3z^{5\over 2}}, \end{split}\end{eqnarray}\end{center}
Next, it is required to substitute $\dot z=Nx$, i.e., $\ddot z=N\dot x+\dot N x$, so that the point Lagrangian \eqref{DA1} may be expressed in the following form,

\begin{eqnarray} \label{DA2}\begin{split} L&= -\frac{3\alpha N x^2}{2\sqrt z}+\frac{9\beta}{\sqrt z}\frac{\dot x^2}{N}+\frac{9N\gamma x^4}{4z^{5\over 2}}+ u\bigg({\dot z\over N}-x \bigg). \end{split}\end{eqnarray}
where $\alpha = \alpha_1+\alpha_2$ and the two linear terms (associated with $R$ and $Q$) get combined. In the above equation \eqref{DA2}, the expression $\big({\dot z\over N}-x = 0\big)$ is treated as a constraint and introduced through the Lagrangian multiplier $u$. The canonical momenta are,
\begin{eqnarray}
p_x &=& {18\beta \dot x\over N\sqrt z}; ~~ p_z={u\over N};~~p_N=0=p_u.
\end{eqnarray}
and the primary Hamiltonian reads as,
\begin{equation}\label{AHp}\begin{split}
   & \mathrm{H}_{p_1}= \frac{3N\alpha x^2}{2\sqrt z}+{u\dot z\over N}+\frac{N\sqrt zp_x^2}{36\beta}-\frac{9\gamma Nx^4}{4z^{5\over 2}}-u\left({\dot z\over N}-x\right).
\end{split}
\end{equation}
\begin{figure}
\begin{minipage}[h]{0.47\textwidth}
\centering
\includegraphics[ width=0.8\textwidth] {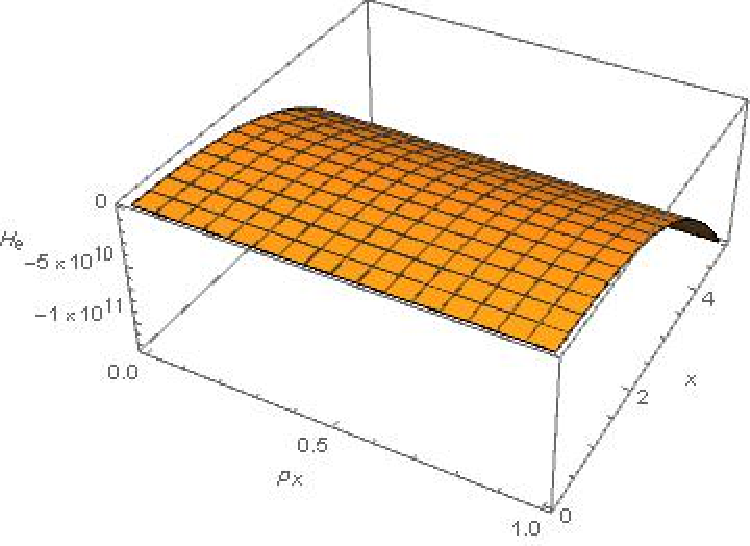}
 \caption{This 3-dimensional plot depicts the variation of $\mathrm{H}_e$ with $p_x$ \& $x$, setting $z = a^2 = 0.001$ and taking $2\alpha={1},~ \beta={1\over 2}, ~~\gamma={4\over 9}$. However, $0 < z \le 1$ does not affect the nature of the plot.}
      \label{fig:1}
   \end{minipage}%
  \hfill
\begin{minipage}[h]{0.47\textwidth}
\centering
\includegraphics[ width=0.8\textwidth] {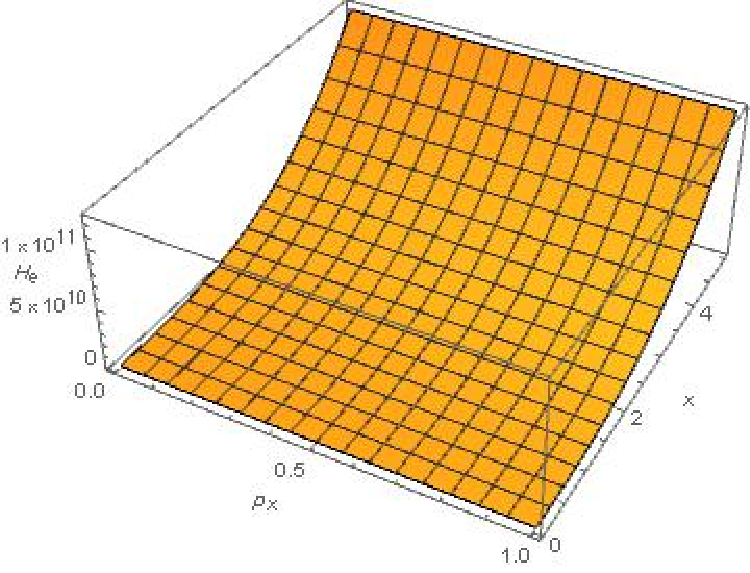}
 \caption{This 3-dimensional plot depicts the variation of $\mathrm{H}_e$ with $p_x$ \& $x$, setting $z = a^2 = 0.001$ and taking $2\alpha={1}, ~\beta={1\over 2}, ~~\gamma=-{4\over 9}$. Note that, $0 < z \le 1$ does not affect the nature of the plot.}
      \label{fig:2}
   \end{minipage}%
   \end{figure}
Now introducing the constraints $\phi_1= N p_z-u \approx 0$ and $\phi_2 = p_u \approx 0$ through the Lagrange multipliers $u_1$ and $u_2$ respectively, we get (Note that, these are second class constraints, since $\{\phi_i, \phi_j\} \ne 0$. Further, since the lapse function $N$ is non-dynamical, so the associated constraint vanishes strongly, ie. $p_N =0$ and therefore it may safely be ignored),
\begin{equation}\label{AHp1}\begin{split}
    \mathrm{H}_{p_1}&=N\bigg[\frac{3\alpha x^2}{2\sqrt z}+ \frac{\sqrt z  p_x^2}{36\beta}-\frac{9\gamma x^4}{4z^{5\over 2}}\bigg]+ux+u_1\big(Np_z-u\big)+u_2p_u.
\end{split}
\end{equation}
It can easily be checked that Poisson brackets $\{x, p_x\} = \{z, p_z\} = \{\phi,p_{\phi}\}=\{u, p_u\} = 1$, hold. Now constraints should remain preserved in time, which are exhibited through the following Poisson brackets
\begin{equation}\begin{split}\label{Apc}
   &\dot\phi_1=\{ \phi_1,\mathrm{H}_{p_1}\}=-u_2-N{\partial\mathrm{ H}_{p_1}\over \partial z}\approx 0\Rightarrow u_2=-N{{\partial \mathrm{H}_{p_1}\over \partial z}};\\& \dot\phi_2= \{\phi_2,\mathrm{H}_{p_1}\}\approx 0 = u_1 - x \Rightarrow u_1=x.\end{split}
\end{equation}
Therefore the primary Hamiltonian is modified to
\begin{equation}\label{AHp2}\begin{split}
    \mathrm{H}_{p_2}&=N\bigg[xp_z+ \frac{3\alpha x^2}{2\sqrt z}+ \frac{\sqrt z  p_x^2}{36\beta}-\frac{9\gamma x^4}{4z^{5\over 2}}\bigg]-Np_u{\partial \mathrm{H}_{p_1}\over \partial z}.
\end{split}
\end{equation}
As the constraint should remain preserved in time in the sense of Dirac, so
\begin{equation}\label{Hp2}
    \dot\phi_1=\{\phi_1,\mathrm{H}_{p2}\}=-N\bigg[{\partial \mathrm{H}_{p_1}\over \partial z}-Np_u{\partial^2\mathrm{H}_{p_1}\over \partial z^2} \bigg]+N{\partial \mathrm{H}_{p_1}\over \partial z}
\approx 0\Rightarrow p_u=0.
\end{equation}
Finally the phase-space structure of the Hamiltonian, being free from constraints reads as,
\begin{equation}\label{AHF}\begin{split}
    \mathrm{H}&=N\bigg[xp_z + \frac{\sqrt z  p_x^2}{36\beta} + \frac{3\alpha x^2}{2\sqrt z}-\frac{9\gamma x^4}{4z^{5\over 2}}\bigg]=N\mathbb{H} = 0,
\end{split}
\end{equation}
since, the Hamiltonian is constrained to vanish. `Diffeomorphic invariance' is thus established. The above Hamiltonian is well behaved as we shall see in next.

\subsection{Canonical quantization \& semiclassical approximation:}
In order to quantize the Hamiltonian (\ref{AHF}) following `standard canonical quantization' scheme, the Hamiltonian is expressed in the following form,
\begin{equation} \label{qh} -{p_z\over \sqrt z}={p_x^2\over 36\beta x} + {3\alpha x\over 2z} -{9\gamma x^3\over 4z^{3}}=\mathrm{H}_e,\end{equation}
where, $\mathrm{H}_e$ is the effective Hamiltonian. The `phase space structure' of the effective Hamiltonian $\mathrm{H}_e$ is depicted through the $3$-dimensional plots $(\mathrm{H}_e ~\mathrm{vs}~ p_x~ \mathrm{vs}~ x)$ in figure-1 for $(\gamma >0)$ and in figure-2 taking $(\gamma<0)$, setting the scale at arbitrary value $z = 0.001$, which corresponds to an early value of the scale factor. Clearly, the `effective Hamiltonian' is well-behaved and the nature of the plot remains unaltered for arbitrary choice of $0 < z = a^2 \le 1$.\\

\noindent
Thus, we finally arrive at the following `modified Wheeler-de-Witt equation',
\begin{eqnarray}\label{WDH}{i\hbar\over \sqrt{z}}\frac{\partial\Psi}{\partial z}=-\frac{\hbar^2}{36\beta x}\bigg{(}\frac{\partial^2}{\partial x^2} +\frac{n}{x}\frac{\partial}{\partial x}\bigg{)}\Psi+\left[ {3\alpha x\over 2z} -{9\gamma x^3\over 4z^{3}}\right]\Psi=\widehat{\mathrm{H}}_e \Psi,\end{eqnarray}
where ${\widehat{\mathrm{H}}}_e$ is the `effective Hamiltonian operator'. Finally, under a change of variable $\sigma=z^{\frac{3}{2}} = a^3$, the above `modified Wheeler-de-Witt equation' takes the form:
\begin{eqnarray}{i\hbar}\frac{\partial\Psi}{\partial \sigma}=-\frac{\hbar^2}{54\beta x}\bigg{(}\frac{\partial^2}{\partial x^2} +\frac{n}{x}\frac{\partial}{\partial x}\bigg{)}\Psi+\left[ {\alpha x\over \sigma^{2\over 3}} -{3\gamma x^3\over 2\sigma^{2}}\right]\Psi= {\widehat{\mathrm{H}}}_e\Psi,\end{eqnarray}
where, the proper volume, $\sigma = a^3$ plays the role of `internal time parameter'. It is noticeable that the effective potential $V_e(x)={3\gamma x^3\over 2\sigma^{2}}$ may be extremized to yield
\begin{eqnarray} x^2 = {2\alpha\over 9\gamma} \sigma^{4\over 3} = {2\alpha\over 9\gamma}z^2.\end{eqnarray}
Since $x = \dot z$, under the choice of the gauge $N = 1$, so upon integration one finds
\begin{eqnarray}\label{DS} z = z_0 e^{\sqrt{2\alpha\over 9\gamma} t}, ~~\mathrm{or}~~ a(t) = a_0 e^{\sqrt{\alpha\over 18\gamma} t},\end{eqnarray}
and de-Sitter solution emerges. This is a promising outcome of the model under consideration. Further, the above `modified Wheeler-de-Witt equation' also looks very much like the standard `Schr\"odinger equation'. The `effective Hamiltonian operator ${\widehat{\mathrm{H}}}_e$' is hermitian under the choice $n = -1$ as referred in \cite{MKA} and correspondingly the continuity equation can be expressed as,
\begin{eqnarray} \frac{\partial\rho}{\partial\sigma}+\nabla. \textbf{J}=0, \end{eqnarray}
where, $\rho=\Psi^* \Psi$ and $\textbf{J}=(\textbf{J}_x, 0, 0)$ are the probability density and current density respectively and
\begin{eqnarray}
\textbf{J}_x &=& \frac{i\hbar}{54\beta x}\big{(}\Psi\Psi^*_{,x}-\Psi^*\Psi_{,x} \big{)}
\end{eqnarray}
All these results are very similar to the previously obtained $f(R,\mathrm{T})$ gravity theory, where $\mathrm{T}$ stands for the the torsion scalar \cite{MKA}, although the action is not exactly the same. To justify the `quantum equation' (\ref{WDH}), we have also studied its behaviour under an appropriate `semiclassical approximation', following the standard Wentzel-Kramers-Brillouin (WKB) approximation assuming $\Psi(x,z) = \Psi_0 e^{{i\over \hbar} S(x,z)}$, $\Psi_0$ being the slowly varying amplitude and expanding $S(x,z)$ in the power series of $\hbar$ as usual \cite{MKA}. The wave function had been found to execute oscillatory behaviour under first order approximation as
\begin{eqnarray}\Psi= \psi_{01} e^{\frac{i}{\hbar}\left[-12\beta \Lambda^3(1\mp\sqrt{1-{\gamma \over 2\beta}})z^{3\over 2}\right]},\end{eqnarray}
where, $\Lambda = \sqrt{2\alpha\over 3\gamma}$. It may be mentioned that under higher-order approximation too, the form of the wave function remains unaltered, while only the pre-factor changes. Therefore the quantum universe smoothly transits to the classical universe and the model is suitable to study inflation in view of the classical field equations, according to `Hartle criteria'. It is noticeable that the parameter $\gamma$ which is the co-efficient of $Q^2$ term determines the degree of oscillation through the constant $\Lambda$. However, $\gamma \le 2\beta$ for ensuring oscillatory behaviour.

\subsection{Inflation:}

Having proved the viability of the action \eqref{A1.0} in the quantum domain, we now proceed to formulate inflationary dynamics under `slow roll approximation'. In this process, we constrain the parameters of the theory and find the `inflationary parameters' and check consistency with the latest released data sets from Planck's mission which constrained inflationary parameters tightly  \cite{Planck1,Planck2,Planck3}. The observational data that we shall compare with our theoretical data is presented in Table-1. \\

`Inflation' is essentially a quantum phenomena, which was initiated sometime between ($10^{-43} - 10^{-26}$) sec., after gravitational sector transits to the classical domain. To be more specific, inflation is a `quantum theory of perturbations' on the top of classical background, which means that the `energy scale' of the background must be much below the Planck scale. The `string theory' swampland also indicates that the energy scale must be rather low for inflation. According to `Hartle criteria', ``if a quantum theory admits a viable semiclassical approximation, then most of the important physics may be extracted from the classical action itself'' \cite{Hartle}. Clearly, the above `semiclassical approximation' is validated if the `energy scale of inflation' is below the `Planck's scale'. Although, inflation may be triggered by higher order curvature invariant term such as $R^2$ (from which a scalar degree of freedom emerges under linearization), also called curvature induced inflation, however, we prefer slow-roll inflation, in which $R^2$ term does not play any role \cite{AKS1,AKS2,AKS3,AKS4}, that would be apparent here too. We therefore consider scalar field ($\phi$) induced slow-roll inflation instead, since `dilatonic scalar field' is a natural outcome under four dimensional compactification of higher dimensional theories formulated to cast quantum gravity theories, such as the string theory, M-theory etc. Hence, incorporating a scalar field in the very early universe, we now write down the two independent field equations, viz., the ${(^0_0)}$ equation of  Einstein and the $\phi$ variation equation, in the Robertson-Walker minisuperspace \eqref{RW}, which are,
\begin{eqnarray}\begin{split}\label{00phi} &\frac{6\alpha\dot a^2}{a^2N^2}+\frac{36\beta}{a^2N^4}\Bigg(2\dot a\dddot a-2\dot a^2\frac{\ddot N}{N}-\ddot a^2-4\dot a\ddot a\frac{\dot N}{N}+2\dot a^2\frac{\ddot a}{a}+5\dot a^2\frac{\dot N^2}{N^2}-2\frac{\dot a^3\dot N}{a N}-3\frac{\dot a^4}{a^2}\Bigg)\\&\hspace{2.5 in}-\frac{108\gamma {\dot a}^4}{N^4a^4}={1\over 2N^2}{\dot\phi}^2+V(\phi)\\&
\ddot\phi +3{\dot a \over Na}\dot\phi +V'=0, \end{split}\end{eqnarray}
where $\alpha= \alpha_1+\alpha_2$, as mentioned before and prime now denotes the derivative with respect to the scalar field $\phi$. However it is crucial to note that, all the coefficients $(\alpha_1, \alpha_2, \beta, \gamma)$ are kept arbitrary from the very beginning. It is well known from the literature that in $f(R)$ theory, the signatures of $\alpha_1$ and $\beta$ must be positive. On the other hand, it has been noticed that in $f(Q)$ theory the energy condition requires $\alpha_2$ to be positive and $\gamma$ has to be negative. It may be mentioned that the pathology of branching appears if both are chosen to be positive \cite{DA2,MKA,conf} and the issue is automatically resolved in view of the energy conditions \cite{DA2,conf}. However, due to the presence of $R^2$ term in $f(R,Q)$ theory, although the issue of branching does not appear, still we consider the same signatures of the coefficients as in $f(Q)$ theory and replace $\gamma$ by $(-\gamma)$.\\

\begin{table}
\begin{minipage}[h]{0.90\textwidth}
      \centering
      \begin{tabular}{|c|c|c|}
     \hline\hline
      $\mathrm{Spectral~Index:}~ n_s$ & $\mathrm{Tensor~to~scalar~ratio:}~r$& ${\mathrm{Number~of~ e-folds:~ N}}$\\
      $\mathrm{It~ shows~up}$ & $\mathrm{BICEP/Keck~(BK18)}$ & $\mathrm{between~Hubble~crossing}$\\
      $\mathrm{zero~ scale ~dependence}$ &$\mathrm{with~BAO~ \&~Planck ~(PR4)}$ &$ \mathrm{at~the~pivot~scale}$\\
      $\mathrm{at}~ 68 \% ~\mathrm{confidence~level}$ & $\mathrm{At}~ 95 \%~\mathrm{confidence~level}$ & $\mathrm{and~the~end~of~inflation}$\\
      \hline
      0.9649 $\pm$ 0.0042 & $<$ 0.032 & 50-60\\
      \hline\hline
      \end{tabular}
     \caption{Observational constraints on inflationary parameters set forth by Planck satellite mission, analyzing $6$-parameter $\Lambda\mathrm{CDM}$ model \cite{Planck1,Planck2,Planck3}.}
      \label{tab:Table1}
      \end{minipage}
      \end{table}
To start with, we first rearrange the ($^0_0$) and the $\phi$ variation equations of Einstein, viz., (\ref{00phi}) respectively for a particular gauge $(N = 1)$ indicating proper time as,
\begin{eqnarray}\label{00H}\begin{split}  &6\alpha{H}^2 + 36\beta{H}^4 \bigg{[}4\bigg{(}1+\frac{\dot {{H}}}{{H}^2}\bigg{)}+ 4\frac{\dot{ {H}}}{{H}^2}\bigg{(}1+\frac{\dot{{H}}}{{H}^2}\bigg{)} +2\bigg{(}\frac{\ddot {{H}}}{{H}^3}-2\frac{{\dot {{H}}}^2}{{H}^4}\bigg{)}-\bigg{(}1+\frac{\dot {{H}}} {{H}^2} \bigg{)}^2-3\bigg{]}\\&\hspace{3.0 in} + {27\over 4}\gamma {H}^4 = V+\frac{\dot\phi^2}{2},\\&
\ddot\phi +3{H}\dot\phi + V' = 0,\end{split} \end{eqnarray}
where $H={\dot a \over a}$ denotes the expansion rate, which is assumed to be slowly varying during inflation. Now, instead of `standard slow roll parameters', we introduce a `hierarchy of Hubble flow parameters' \cite{AKS1, AKS2, SR}, since it turned out to be much worthy to handle higher order theories. Firstly, a set of `horizon flow functions' (the behaviour of Hubble distance during inflation) describes the `background evolution' of the theory under consideration stemming from,
\begin{eqnarray} \label{dh}\epsilon_0=\frac{d_{{H}}}{d_{{H}_i}},~ \text{where}~~ d_{{H}}={H}^{-1},\end{eqnarray}
where, $d_{{H}} = {H}^{-1}$ is the `Hubble distance or the horizon' in our chosen units. We use suffix $i$ to denote the era at which inflation commenced. Now the hierarchy of functions is defined methodically as,
\begin{eqnarray} \label{el} \epsilon_{l+1}=\frac{d\ln|\epsilon_l|}{d \mathrm{N}},~~l\geq 0. \end{eqnarray}
Considering the definition of the number of e-folds $\mathrm{{N}}=\ln{\left(\frac{a}{a_i}\right)}$, implying ${\dot {\mathrm N}}={H}$, it is possible to compute the `logarithmic change of Hubble distance per e-fold expansion ${\mathrm{N}}$', which is the `first slow-roll parameter', as $\epsilon_1=\frac{d\ln{d_H}}{d{\mathrm{N}}} = {{\dot d_{H}}\over \mathrm{N}d_{H}} ={{\dot d_{H}}\over \mathrm{N}{H}^{-1}}=\dot{d_H}=-\frac{\dot H}{H^2}$.
This makes sure that the Hubble parameter remains almost constant during inflation, i.e., as long as $\epsilon_1\ll 1$. The above `hierarchy' also allows to determine $\epsilon_2=\frac{d\ln{\epsilon_1}}{d{\mathrm{N}}}=\frac{1}{{H}}\frac{\dot\epsilon_1}{\epsilon_1},$ which implies $\epsilon_1\epsilon_2 = {\dot \epsilon_1\over H} = d_{{H}} \ddot{d_{{H}}} = -\frac{1}{{H}^2}\left(\frac{\ddot {{H}}}{{H}}-2\frac{\dot {{H}}^2}{{H}^2}\right)$. `Higher slow-roll parameters' may be computed in the same manner as well. Equation (\ref{el}) essentially defines a flow in space with cosmic time being the evolution parameter, which is described by the equation of motion
\begin{eqnarray}\label{el1}\epsilon_0\dot\epsilon_l-\frac{1}{d_{{H}_i}}\epsilon_l\epsilon_{l+1}=0,~~~~l\geq 0.\end{eqnarray}
Having established the hierarchy of the `Hubble flow parameters', we now turn our attention back to the field equations. In view of the `slow-roll parameters', equations (\ref{00H}) may therefore be expressed as,
\begin{eqnarray} \label{hir1}\begin{split}& -6\alpha {H}^2-36\beta {H}^4\left[3\big(1-\epsilon_1\big)^2-2\big(1+\epsilon_1 \epsilon_2 \big)-1\right] - {108}\gamma {H}^4+\Big(\frac{\dot\phi^2}{2}+V\Big)=0,\\& \ddot\phi +3{H}\dot\phi=-V'\end{split}\end{eqnarray}
respectively, which may therefore be approximated using the `slow roll hierarchy' ($\epsilon_1 \ll 1$, $\epsilon_1\epsilon_2 \ll 1$) to,
\begin{eqnarray} \begin{split}\label{com}& 6\alpha{H}^2+108\gamma H^4 = \frac{\dot\phi^2}{2} + V,\\& \ddot\phi +3{H}\dot\phi + V'=0.\end{split}\end{eqnarray}
Finally, imposing the `standard slow roll conditions', viz. $\dot\phi^2 \ll V(\phi)$ and $|\ddot \phi| \ll 3{H}|\dot \phi|$, the above equations are further simplified to,
\begin{eqnarray} \label{hir11}\begin{split}& 6\alpha{H}^2+108\gamma H^4 =  V, \hspace{2cm} 3{H}\dot\phi + V'=0. \end{split}\end{eqnarray}
Note that, the contribution of higher order term due to $R^2$ does not appear under `slow roll approximation', but the higher-degree term due to $Q^2$ is very much present in the form of $H^4$. To proceed further, we solve for $H^2$ to arrive at (considering the positive sign for obvious reason),
\begin {align} H^2 = \frac{\left[-\alpha+ \sqrt{\alpha^2+12\gamma V(\phi)}\right]}{36\gamma}.\label{soln}\end{align}
Further, combining equations of (\ref{hir11}) the slow-roll parameters can be computed as,
\begin {align}\label{SR2}\begin{split} \epsilon={1\over 16\pi G}\bigg({V'\over V} \bigg)^2 = - {6\alpha\dot {H}\over {H}^2}\bigg[\frac{\alpha +36\gamma H^2}{\big(\alpha +18\gamma H^2\big)^2}\bigg];\hspace{0.3 in}\eta = 2 \alpha \left({V''(\phi)\over V(\phi)}\right),\end{split}\end {align}
where, the relation of $\epsilon$ exhibits the fact that the `potential slow roll parameter' is equal to the `Hubble slow roll parameter' only when $\gamma = 0$. Also, in view of equations \eqref{hir11}, one finds $\frac{H}{\dot\phi}=-\frac{\left[-\alpha+\sqrt{\alpha^2+12\gamma V(\phi)}\right]}{12\gamma V'(\phi)}$, which when used, the `number of e-folds' may be computed as,
\begin{align}\label{Nphi2} {\mathrm{N}}(\phi)\simeq \int_{t_i}^{t_f}{H}dt=\int_{\phi_i}^{\phi_f}{{H}\over {\dot\phi}}d\phi= \int_{\phi_f}^{\phi_i}\frac{\left[-\alpha+\sqrt{\alpha^2+12\gamma V(\phi)}\right]}{12\gamma V'(\phi)}d\phi.\end{align}
At this stage, we need to choose a form of the potential.\\

\noindent
\textbf{Case-1:}
Out of a possible host of potentials, constraints from Planck \cite{Planck1,Planck2} rule out all simple integer power law potentials for single field inflation, strongly excluding the standard quadratic ($\phi^2$), $\phi$ and $\phi^{2\over 3}$ at $5\sigma$, while favouring hilltop-like potentials with $V''(\phi) <0$. Here therefore as an example, we first choose one such potential,
\begin{eqnarray}\label{V1} V(\phi)= V_0-{V_1\over \phi},\end{eqnarray}
since it remains almost flat initially for large value of the scalar field $\phi$ and admits slow-roll, with $V'' < 0$ for $V_1 > 0$. The expressions of $\epsilon, ~\eta$ \eqref{SR2} and $\mathrm{N}$ \eqref{Nphi2} may then be found as,
\begin {align}\label{para2}\begin {split} &\epsilon =  \frac{\alpha}{\phi^4\left({V_0\over V_1}-{1\over \phi}\right)^2};\hspace{1 in}\eta =-\frac{4\alpha}{\phi^2(\frac{V_0}{V_1}\phi-1)};\\& \mathrm{N} = \int_{\phi_f}^{\phi_i}\frac{3\phi^2\left[-3\alpha +\sqrt{9\alpha^2+108\gamma V_1\left(\frac{V_0}{V_1}-\frac{1}{\phi}\right)}\right]}{108\gamma V_1}d\phi.\end{split}\end{align}
The data set varying ${V_0 \over V_1}$ between $4.1M_P^{-1} \leq {V_0 \over V_1}\leq  4.8 M_P^{-1}$ presented in Table-2 reveals the fact that the agreement with observation (Table-1) is outstanding in respect of the `tensor to scalar ratio $r$' and the `spectral index $n_s$'. The variation of $n_s$ versus $r$ is portrayed in Fig-3. The `number of e-folds' lie within the range $40 \leq \mathrm{N} \leq 47$, which is although a little low, but possibly is sufficient to solve the `horizon and flatness problems'. Particularly, the last three data are appreciably good. It is worth remembering the fact that the `tensor to scalar ratio' had been constrained from ``$r < 0.32$ to $r < 0.055$'' in only one decade by Planck satellite mission and may turn out to be even small in the future experiments. The current model can sustain further constraints on $r$. It may be stated that $\alpha = \alpha_1 + \alpha_2 = {1\over 16\pi G} = {M_P^2\over 2} = 0.5 M_P^2$ is the choice made by Einstein to recover Poisson equation under weak field approximation, which we have chosen here and therefore should not be treated as a constraint of the theory. \\

\begin{table}
   \begin{minipage}[h]{0.48\textwidth}
      \centering
      \begin{tabular}{|c|c|c|c|c|}
     \hline\hline
      $\phi_f~(M_P)$ &${V_0\over  V_1}~(M_P^{-1})$ & $n_s$ & $r$ & $ {\mathrm{N}}$\\
      \hline
        0.55478 &4.1 &  0.9658 & 0.00532 & 40\\
        0.54628 &4.2 &  0.9667 & 0.00505 & 41\\
        0.53814 &4.3 &  0.9676 & 0.00480 & 42\\
        0.53031 &4.4 &  0.9684 & 0.00457 & 43\\
        0.52279 &4.5 &  0.9692 & 0.00435 & 44\\
        0.51555 &4.6 &  0.9699 & 0.00415 & 45\\
        0.50858 &4.7 &  0.9707 & 0.00396 & 46\\
        0.50187 &4.8 &  0.9713 & 0.00379 & 47\\
    \hline\hline
    \end{tabular}
     \caption{Theoretical data set of the inflationary parameters, for the potential $V = V_0- {V_1\over \phi}$ (Case-1 \eqref{para2}), under the variation of ${V_0 \over V_1}$, taking $\phi_i=3.2 M_P$, $108{\gamma V_1}=0.005 M_P^{5}$,~$\alpha=0.5 M_P^2$.}
      \label{tab:Table2}
   \end{minipage}%
   \hfill
   \begin{minipage}[h]{0.45\textwidth}
      \centering
      \begin{tabular}{|c|c|c|c|c|}
     \hline\hline
      $\phi_f$ in $M_P$ &${V_0\over  V_1}$  & $n_s$ & $r$ & $ {\mathrm{N}}$\\
      \hline
        0.05024 &1.194 &  0.9601 & 0.02215 & 58  \\
        0.04466 &1.196 &  0.9602 & 0.02206 & 59 \\
        0.03909 &1.198 &  0.9603 & 0.02198 & 59 \\
        0.03353 &1.200 &  0.9604 & 0.02189 & 59  \\
        0.02244 &1.204 &  0.9606 & 0.02172 & 59  \\
        0.01133 &1.208 &  0.9608 & 0.02155 & 59  \\
        0.00587 &1.210 &  0.9609 & 0.02147 & 60  \\
        0.00036 &1.212 &  0.9610 & 0.02138 & 60  \\
    \hline\hline
    \end{tabular}
     \caption{Theoretical data set of the inflationary parameters for the potential $V= V_0-{V_1\exp(-b\phi)}$ (Case-2 \eqref{para3}), varying ${V_0 \over V_1}$, taking $\phi_i=5.75 M_P$, ${\gamma V_1}=0.130 M_P^{4}$,~$\alpha=0.5 M_P^2,~b=0.3M_P^{-1}$.}
      \label{tab:Table3}
   \end{minipage}%
  \end{table}

 \begin{figure}
\begin{minipage}[h]{0.42\textwidth}
\centering
\includegraphics[ width=1.0\textwidth] {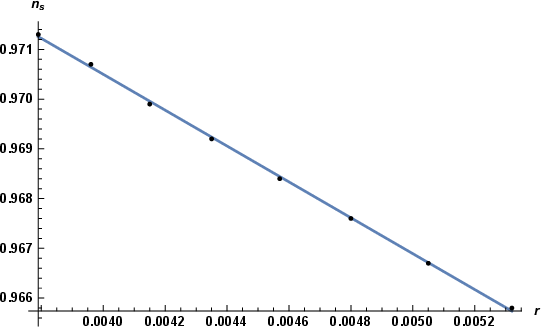}
 \caption{Variation of $n_s$ with $r$ for the potential $[V=V_0-{V_1\over \phi}]$, Case-1.}
      \label{fig:3}
   \end{minipage}%
   \hfill
\begin{minipage}[h]{0.40\textwidth}
\centering
\includegraphics[width=1.0\textwidth] {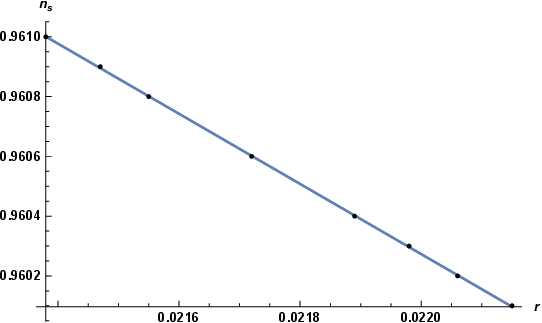}
 \caption{Variation of $n_s$ with $r$ for the potential $[V=V_0-{V_1\exp(-b\phi)}]$, Case-2.}
      \label{fig:4}
   \end{minipage}
   \end{figure}

Next, we compute the `energy scale of inflation' considering the relation \eqref{soln}, choosing one of the data presented in Table 1, viz., ($\mathrm{N} = 45$, for which $\phi_i = 3.2M_P,~ \alpha=0.5M_P^2, ~{{V_0\over V_1}=4.6 M_P^{-1}},~108\gamma V_1= 0.005M_P^5$). Correspondingly, we find,
\begin{align}\label{ES1}{H_*}^2 = {0.00712\over 108\gamma},\end{align}
which is still arbitrary, since $\gamma$ has not been fixed as yet. Now, the ``energy scale of inflation for a single scalar field model'' as already considered earlier, is given by the following relation \cite{Wands},
\begin{align}\label{ES2}{H_*} = 8\times 10^{13}\sqrt{r\over 0.2}~GeV=1.15\times 10^{13}~GeV =0.4694\times 10^{-5} M_p.\end{align}
Note that, to compute the numerical value of $H_*$ appearing in \eqref{ES2} we take into account the tensor-to-scalar ratio $r = 0.00415$ which corresponds to $\mathrm{N} = 45$, under consideration (see Table 1). Other way round for a cross check, let us consider that inflation lasted for $\Delta t \approx 10^{-35}~s$, if it is assumed to occurs between $(10^{-42}$ to $10^{-35})~s$. Then for the same value of $\mathrm{N} = 45$ (which is precisely $\Delta \mathrm{N}$), $H \approx {45\over 10^{-35}}~s^{-1}$. However, in our chosen system of unit, $1 s^{-1} \approx {1\over 3.65 \times 10^{42}}~M_P$. Hence, $H = H_* \approx 0.123 \times 10^{-5}~M_P$. Thus, to arrive at this scale of inflation, we need to constrain $\gamma \approx 2.99 \times 10^{6}$. Once the parameter $\gamma$ is fixed from physical consideration (sub-Planckian scale of inflation), the parameters $V_1$ and $V_0$ are found as well,
\begin{align} \label{V2}  V_1 \approx 1.55\times 10^{-11}M_P^5, \hspace{0.2 in} V_0 = 4.6 V_1\approx 7.13\times 10^{-11}M_P^4.\end{align}
Finally, it is now required to check if the model admits `graceful exit from inflation'. We therefore recall the first equation of \eqref{com}, which in view of the above form of the potential, $V(\phi)=V_0-{V_1\over \phi}$, is expressed as,
\begin{align}\label{H1}-\frac{\gamma H^4}{V_1} - \frac{6\alpha H^2}{V_1}+\left[{\dot\phi^2\over 2V_1}+\left({V_0\over V_1}-{1\over {\phi}}\right)\right]=0.\end{align}
Initially, during inflation all the terms in \eqref{H1} are more-or-less of the same order of magnitude, since $H$ is slowly varying. Nonetheless as inflation halts, the `Hubble parameter' falls of sharply and so one can neglect both the terms ${\gamma H^4\over V_1}$ and ${\alpha H^2\over V_1}$ appearing in the left hand side of equation \eqref{H1} without sacrificing generality to arrive at,
\begin{align}{\dot{\phi}}^2=-2\left[V_0-{V_1\over \phi}\right] .\label{osc2}\end{align}
Taking into account, $\phi_f=0.5156M_P$, $V_0= 7.13\times 10^{-11}M_P^4$ and $V_1= 1.55\times 10^{-11}M_P^5$ from Table-1, the `oscillatory behavior' of the scalar field at the end of inflation is exhibited in the following equation,
\begin{align}\phi=\exp({i\omega t}),\end{align}
provided, $\omega \approx  1.762\times10^{-5}M_P$. Therefore, `graceful exit from inflation' is manifest. In a nutshell, the potential \eqref{V1} is well-suited in the very early stage of cosmic evolution.\\

\noindent
\textbf{Case-2}:
Next, for the sake of generality, we consider a decaying exponential potential in the following form,
\begin{eqnarray}\label{V2} V(\phi)= V_0-{V_1\exp(-b\phi)}.\end{eqnarray}
The above form of the potential has extensively been used to study inflation because, not only it admits slow roll when $\phi$ is large enough, but also as the scalar field dwindles as a consequence of particle production at the end of inflation, the potential becomes almost a constant and acts as an effective cosmological constant. Also, $V''(\phi) < 0$, for $V_1 > 0$.\\

In view of the above form of potential \eqref{V2}, the expressions of $\epsilon, ~\eta$ \eqref{SR2} and $\mathrm{N}$ \eqref{Nphi2} are found as,
\begin {align}\label{para3}\begin {split} &\epsilon =  \frac{\alpha b^2\exp({-2b\phi})}{\left[{V_0\over V_1}-\exp({-b\phi})\right]^2};\hspace{1 in}\eta =-\frac{2\alpha b^2\exp({-b\phi})}{\left[{V_0\over V_1}-\exp({-b\phi})\right]};\\& \mathrm{N}=\int_{\phi_f}^{\phi_i}\frac{\left[-\alpha +\sqrt{\alpha^2+12\gamma V_1\left(\frac{V_0}{V_1}-\exp(-b\phi)\right)}\right]}{12\gamma V_1b\exp({-b\phi})}d\phi.\end{split}\end{align}
The data set varying ${V_0 \over V_1}$ between $1.194 \leq {V_0 \over V_1}\leq  1.210 $ presented in Table-3 reveals the fact that the agreement with observation (Table-1) is outstanding in respect of the `tensor to scalar ratio $r$' and the `spectral index $n_s$'. Additionally, the `number of e-folds' lie within the range $58 \leq \mathrm{N} \leq 60$, which is sufficient to solve the `horizon and the flatness problems'. The variation of $n_s$ with $r$ is depicted in figure-4.\\

Next as before, we compute the `energy scale of inflation' considering the relation \eqref{soln}, choosing one of the data presented in Table-3, viz., ($\mathrm{N} = 60$, for which $\phi_i = 5.75M_P,~ \alpha=0.5M_P^2, ~{V_0 \over V_1}=1.210,~\gamma V_1= 0.130M_P^4$). Correspondingly, we find,
\begin{align}\label{ES3}{H_*}^2 = {0.864\over 36\gamma},\end{align}
which is still arbitrary, since $\gamma$ has not been fixed as yet. Now, the energy scale of inflation for a single scalar field model as already considered earlier, is given by the following relation \cite{Wands},
\begin{align}\label{ES4}{H_*} = 8\times 10^{13}\sqrt{r\over 0.2}~GeV=2.62\times 10^{13}~GeV =1.07\times 10^{-5} M_p,\end{align}
where, we have taken $r = .02147$, which corresponds to $\mathrm{N}=60$ as appears in Table-2. Thus, to arrive at that scale of inflation, we need to constrain $\gamma \approx 2.99 \times 10^{6}$. After fixing the parameter $\gamma$ from physical consideration (sub-Planckian scale of inflation), the parameters $V_1$ and $V_0$ are found as well, which are
\begin{align} \label{2V2}  V_1 \approx 6.18 \times 10^{-10}M_P^4, \hspace{0.2 in} V_0 = 1.210 V_1\approx 7.48\times 10^{-10}M_P^4.\end{align}
Finally, to check if the model admits `graceful exit from inflation', we proceed as before taking into account $\phi_f = 0.00587M_P$, $V_0 = 7.48\times 10^{-10}M_P^4$ and $V_1 = 6.18\times 10^{-10}M_P^4$ from Table-2. Consequently, we find that the `oscillatory behavior' of the `scalar field' at the `end of inflation' is exhibited by the following equation,
\begin{align}\phi=\exp({i\omega t}),\end{align}
provided, $\omega \approx  2.758\times10^{-3}M_P$. Therefore, `graceful exit from inflation' is manifest.\\

\textbf{Summary}: Summarily, the drawbacks of GSTG led us to consider an even generalized theory in the form of $f(R, Q)$. We have constructed the phase-space structure for a potentially viable form of $f(R, Q)$, quantized the theory which admits a Schr\"odinger-like modified Wheeler-deWitt equation rendering quantum-mechanical probabilistic interpretation, and an oscillatory semi-classical wave function that allows to study inflation in view of the classical field equations. The inflationary parameters agree appreciably with the latest released Planck's data, at the sub-Planckian scale of inflation, which admits graceful exit giving way to the radiation dominated era. It is note-worthy, that the exponentially decaying  potential (Case-2) is more feasible, since not only it is flat to admit slow-roll and ends up with an effective cosmological constant upon particle production, but also the agreement with Planck's data is astounding. It is noticeable that the parameter $\gamma$ (the coefficient of $Q^2$ term) plays a major role in these regard, which requires to be constrained identically in both the cases.

\section{Concluding remarks}

`Generalized teleparallel gravity theories' as an alternative to the dark energy and the modified $f(R)$ theory of gravity, were proposed and manifested its excellence to combat cosmic puzzle. However, in the last fifteen years or so, intense study on both the generalized metric teleparallel gravity (GMTG) and generalized symmetric teleparallel gravity theory (GMTG) reveled that both the theories suffer from some serious pathologies. These include `failure of linear perturbation' for maximally symmetric space-time', `strong coupling issue', `appearance of ghost degree of freedom', `loosing diffeomorphic invariance' etc. Recently, it is also revealed that the two of the three connections associated with maximally symmetric flat Robertson-Walker metric, do not admit GSTG in the form of $f(Q)$ theory. The first connection, on the other hand, although admits GSTG and `diffeomorphic invariance' in the absence of the shift vector, the Hamiltonian is not manageable and amenable to canonical quantization. Here we exhibit the fact that the fourth connection associated with non-flat ($k \ne 0$) RW metric although admits GSTG in the form $f(Q)$, and diffeomorphic invariance is admissible as well in the absence of the shift vector, the Hamiltonian is again eerie and not amenable to canonical quantization. These facts surely go against acceptability of the teleparallel theories and modification in some form is required. Under such circumstances, we assume that both `the curvature and non-metricity' are simultaneously responsible to govern gravity and consider a ``generalized $f(R,Q)$ theory of gravity'', $R$ being constructed from the `Levi-Civita connection', while $Q$ stands for the non-metricity scalar. This is justified, since whenever a metric is chosen, one can find the Levi-Civita connections and consequently the Ricci tensor, while the difference of two connections (affine connection and the Levi-Civita connection) results in a tensor, which is the disformation tensor in the present case. \\

We have chosen a realistic model that is viable in the early universe and produces a reasonably viable Hamiltonian. The Hamiltonian is well behaved, and the extremization of the effective potential leads to de-Sitter universe. Canonical quantization shows up the Schr\"odinger-like equation in which the proper volume acts as the effective time parameter. Standard probabilistic interpretation and continuity equation follow under a particular choice of the `operator ordering index $n = -1$'. The `semiclassical approximation' leads to an oscillatory wave-function which meets Hartle criteria. We associate a `scalar field to drive inflation' and make two admissible choices of the potential, both of which remain almost flat in the `early universe' when the `scalar field' is large enough and admits $V(\phi)'' < 0$. Under appropriate slow roll approximation it is found that the $R^2$ term actually does not contribute while the $Q^2$ term contributes largely. We have constrained the parameters of the theory so that the inflationary parameters agree with the last data released from the Planck mission before it ceases to work. It is found that the exponentially decaying potential (Case-2) is better suited. It may also be mentioned that as universe expands, both the $R^2$ as well as $Q^2$ terms become subdominant and a Friedmann-like radiation era and early matter dominated era emerge automatically. To explain late-time comic acceleration, additional term is required in the form $Q^n,~ n < {1\over 2}$ and possibly negative, which dominates only at the late-stage of cosmic evolution. This has been explored earlier by several authors. To the best of our knowledge, $f(R,Q)$ theory in the form chosen here is free from all pathologies.

\section*{Author Contributions}
The problem was suggested by A.K. Sanyal, the formulation was carried out by D. Saha and checked by A.K. Sanyal. The draft was primarily written by D. Saha, and A.K. Sanyal modified and prepared the final version. All authors have read and agreed to the published version of the manuscript.

\section*{Funding Statement}
This research received no external funding.
\section*{Conflict of Interest}

The authors declare no conflict of interest.
\section*{Competing Interest}
The authors declare no competing interest.

\section*{Institutional Review Board Statement}
 Not applicable.

\section*{Data Availability}
Data are contained within the article.

\section{Appendix}
This appendix aims at exhibiting the role of the total derivative terms appearing in the action and the reason for modifying ``Dirac-Bergmann constrained analysis" to formulate the phase space structure of higher-order theories of gravity. To be more specific, we consider the original ``Dirac-Bergmann constrained analysis", in which total derivative terms are not taken care of. Even more, in this formalism, instead of $h_{ij} = a^2 \delta_{ij} = z \delta_{ij}$, the scale factor is treated as the basic variable. However, for the sake of comparison, we treat $h_{ij}$ as the basic variable instead.
\subsection{Dirac-Bergmann constraint analysis:}
In order to perform Dirac-Bergmann constraint analysis, it is necessary to choose an additional variable $x ={{\dot z} \over N} = -2K_{ij}$ and express the action (\ref{A1.0}) in terms of $z$ and $x$ as,
\begin{eqnarray}\begin{split}\label{ADA} A &= \int\left[3\alpha_1\sqrt z\dot x+\frac{9\beta{\dot x}^2}{N\sqrt z}-\frac{3N\alpha_2x^2}{2\sqrt z}+{9N\gamma x^4\over 4z^{5\over 2}}\right]dt\end{split}\end{eqnarray}
Since both $p_z = 0 = p_N$, the Lagrangian corresponding to action \eqref{ADA} becomes singular and thus, it is required to follow Dirac's constraint analysis, to cast the action in the canonical form. We therefore introduce the constraint ${{\dot z }\over N} -x = 0$ through Lagrange multiplier  in action \eqref{ADA}, so that the point Lagrangian now reads as,
\begin{eqnarray} L= \left[3\alpha_1\sqrt z\dot x+\frac{9\beta{\dot x}^2}{N\sqrt z}-\frac{3N\alpha_2x^2}{2\sqrt z}+{9N\gamma x^4\over 4z^{5\over 2}}+u\left({\dot z \over N} -x\right)\right]\end{eqnarray}
The canonical momenta are,
\begin{eqnarray}
  P_x &=& 3\alpha_1\sqrt z+{18\beta \dot x\over N\sqrt z}; ~~ P_z={u\over N};~~P_N=0=P_u.
\end{eqnarray}
Clearly, three primary constraints are present involving Lagrange multiplier or its conjugate viz, $\phi_1=NP_z-u,~\phi_2=P_u \approx 0$, and, $\phi_3 = P_N \approx 0$. However, since the lapse function $N$ is non-dynamical, so the associated constraint $\phi_3$ vanishes strongly. The remaining two are second class constraints, as $\{\phi_1, \phi_2\} \neq 0$.  Therefore, after substituting the remaining first two constraints ($\phi_1,\phi_2$) through the lagrangian multiplier $u_1, u_2$ respectively, the primary Hamiltonian $(H_{p1})$ can be constructed. As the constraints must vanish weakly in the sense of Dirac, therefore analyzing $\dot \phi_1=\{\phi_1, H_{p1}\}\approx0$ and $\dot \phi_2 =\{\phi_2, H_{p1}\}\approx 0$, one can compute $u_1, \& u_2$ respectively and thus imposing these conditions, $H_{p1}$ is modified by the secondary Hamiltonian $H_{p2}$. Again computation of $\dot \phi_1=\{\phi_1, H_{p2}\}\approx0$ results in $P_u = 0$. Thus the Hamiltonian is free from constraints and finally reads as,
\begin{eqnarray}\label{H} \mathrm{H} = N\left[x P_z+\frac{\sqrt z {P_x}^2}{36\beta}-\frac{\alpha_1 z P_x}{6\beta}+\frac{\alpha_1^2z^{3\over 2}}{4\beta}+\frac{3\alpha_2x^2}{2\sqrt z} -\frac{9\gamma x^4}{4z^{5\over 2}}\right]=N\mathbb{H} = 0.\end{eqnarray}

Although the two (\ref{AHF}) and (\ref{H}) exactly match under the following set of canonical transformations,

\[P_z \rightarrow p_z + {3\alpha_1 x\over 2\sqrt z},~~z \rightarrow z,\]
\[P_x \rightarrow p_x+3\alpha_1\sqrt z,~~ x \rightarrow x,\]

and apparently there is no contradiction between the two Hamiltonians \eqref{AHF} and \eqref{H}. Note the essential difference: linear term in the momentum ($zp_x$), which is very much present in (\ref{H}), remains absent from the Hamiltonian \eqref{AHF}. It is not a surprise that both the Hamiltonians \eqref{AHF} and \eqref{H} give unique field equations, since at the classical level the divergent terms are unimportant. However, the classical canonical transformation cannot be carried out in the quantum domain due to non-linearity. Note that the additional linear term $z p_x$ in particular, carries unwanted features in the quantum domain. For example, upon canonical quantization the `modified Wheeler-de-Witt equation' now is expressed as.
\begin{eqnarray}\label{MWD}{i\hbar\over \sqrt{z}}\frac{\partial\Psi}{\partial z}=-\frac{\hbar^2}{36\beta x}\bigg{(}\frac{\partial^2}{\partial x^2}
+\frac{n}{x}\frac{\partial}{\partial x}\bigg{)}\Psi+ {i\hbar \alpha_1\sqrt z\over 6\beta x}\left({\partial \Psi\over \partial x}\right)
+\left[ {\alpha_1\sqrt z\over 4\beta x} + {3\alpha_2 x\over 2z} -{9\gamma x^3\over 4z^{3}}\right]\Psi.\end{eqnarray}
Surely, the effective potential has been modified and the beautiful feature such as the de-Sitter solution \eqref{DS} upon extremization, disappears. However, this is not a major issue. More importantly, the additional second linear term $\left[i\hbar {\partial \Psi\over \partial x}\right]$ on the right hand side desists from constructing a viable quantum dynamical equation. This is the reason for considering ``Modified Dirac-Bergmann constrained analysis".\\
\end{document}